%% file: DEAP_ERBgModel_and_K42_arxive.tex
\documentclass[%
 reprint,
superscriptaddress,
nofootinbib,
 amsmath,amssymb,
 aps,
 prd,
]{revtex4-1}

\usepackage{graphicx}
\usepackage{dcolumn}
\usepackage{bm}

\usepackage{color}
\usepackage{units}
\usepackage{upgreek}          
\usepackage{ifthen}
\usepackage{booktabs}

\usepackage[symbol*]{footmisc}
\DefineFNsymbolsTM{otherfnsymbols}{%
  \textdagger    \dagger
  \textasteriskcentered *
  \textsection   \dagger
  \textbardbl    \|%
  \textparagraph \mathparagraph
  \textbullet \circ
  \textdaggerdbl \ddagger
}%
\setfnsymbol{otherfnsymbols}

\input{deap-abbreviations.tex}
\begin{document}
\include{commands}
\preprint{APS/123-QED}

\title{Electromagnetic Backgrounds and Potassium-42 Activity\\in the DEAP-3600 Dark Matter Detector}

\include{er_authors}

\date{\today}

\begin{abstract}
The DEAP-3600 experiment is searching for WIMP dark matter with a 3.3
tonne single phase liquid argon (LAr) target, located 2.1~km underground
at SNOLAB.  The experimental signature of dark matter interactions is
keV-scale \nuc{Ar}{40} nuclear recoils (NR) producing 128 nm LAr
scintillation photons observed by PMTs. 
The largest backgrounds in DEAP-3600 are electronic recoils (ER)
induced by $\beta$ and \grays\ originating from internal and external
radioactivity in the detector material. 
A background model of the ER interactions in DEAP-3600 was developed and is described in this work.
The model is based on several components which are
expected from radioisotopes in the LAr, from {\it ex-situ} material assay measurements,
and from dedicated independent {\it in-situ}
analyses. 
This prior information is used in a Bayesian fit of the ER
components to a 247.2~d dataset to model the
radioactivity in the surrounding detector materials.  

While excellent discrimination between ERs and NRs is reached with pulse shape discrimination, 
detailed knowledge of the ER background and activity of detector components, sets valuable constraints on other key types of backgrounds in the detector: neutrons and alphas.

In addition, the activity of \nuc{Ar}{42} in LAr in DEAP-3600 is determined by measuring the daughter decay of \nuc{K}{42}. 
This cosmogenically activated trace isotope is a relevant background at higher energies for other rare event searches using atmospheric argon e.g.\ DarkSide-20k, GERDA or LEGEND. 
The specific activity of \nuc{Ar}{42} in the atmosphere is found to be $40.4 \pm 5.9$~$\mu$Bq/kg of argon.

\end{abstract}

\pacs{Valid PACS appear here}
\maketitle


\section{\label{sec:intro}Introduction}

\input{intro}

\section{\label{sec:dataselection}Data Selection}

\subsection{Data Set and Cuts}

\input{datasetcuts}

\subsection{Energy Calibration}

\input{calibration}

\section{\label{sec:backgroundsources}Background Sources}

\input{backgroundsources}

\section{\label{sec:backgrounddecomposition}Fitting Methodology}

\input{decomposition}

\section{\label{sec:results}Results}

\input{results}

\section{\label{sec:externalgammaverification}Verification With External gamma ray Sources}

\input{externalgammaverification}

\section{\label{sec:Ar42K42}Specific Activity of \nuc{Ar}{42} and \nuc{K}{42}}

\input{Ar42K42}

\section{Conclusion}

\input{conclusion}

\begin{acknowledgements}

  We thank the Natural Sciences
  and Engineering Research Council of Canada, the Canadian Foundation
  for Innovation (CFI), the Ontario Ministry of Research and Innovation (MRI), and
  Alberta Advanced Education and Technology (ASRIP), Queen's
  University, the University of Alberta, Carleton University, the
  Canada First Research Excellence Fund, the Arthur B.~McDonald
  Canadian Astroparticle Research Institute, DGAPA-UNAM (PAPIIT
  No. IA100118) and Consejo Nacional de Cienca y Tecnolog\'ia
  (CONACyT, Mexico, Grants No. 252167 and A1-S-8960), the European
  Research Council Project (ERC StG 279980), the UK Science and
  Technology Facilities Council (STFC ST/K002570/1 and ST/R002908/1),
  the Leverhulme Trust (ECF-20130496),  the Spanish Ministry of Science, Innovation and Universities (FPA2017-82647-P grant and MDM-2015-0509). Studentship support from the Rutherford
  Appleton Laboratory Particle Physics Division, STFC and SEPNet
  PhD is achknowledged.  We would like to thank SNOLAB and its
  staff for support through underground space, logistical, and
  technical services. SNOLAB operations are supported by the CFI and Province of Ontario MRI, with underground access provided by Vale at
  the Creighton mine site. We thank Vale for support in shipping the
  acrylic vessel underground.  We gratefully acknowledge the support
  of Compute Canada, Calcul Qu\'ebec, the Center for Advanced
  Computing at Queen's University, and the Computation Center for Particle and Astrophysics (C2PAP) at the Leibniz Supercomputer Center (LRZ) for providing the  computing
  resources required to undertake this work.

\end{acknowledgements}

\bibliography{apssamp}

\clearpage

\section{Appendix}

\input{appendix}

\end{document}

%% file: commands.tex
\definecolor{brown}{RGB}{139,69,19}

\newcommand{\nuc}[2]{$^{#2}\rm #1$}
\newcommand{\nucbf}[2]{$^{\bf #2} \rm \bf#1$}

\newcommand{\bb}[1]{$\rm #1\nu \beta \beta$}
\newcommand{\bbm}[1]{$\rm #1\nu \beta^- \beta^-$}
\newcommand{\bbp}[1]{$\rm #1\nu \beta^+ \beta^+$}
\newcommand{\bbe}[1]{$\rm #1\nu \rm ECEC$}
\newcommand{\bbep}[1]{$\rm #1\nu \rm EC \beta^+$}

\newcommand{\largeGERDA}{{LArGe}}
\newcommand{\PI}{\mbox{\textsc{Phase\,I}}}
\newcommand{\PIa}{\mbox{\textsc{Phase\,Ia}}}
\newcommand{\PIb}{\mbox{\textsc{Phase\,Ib}}}
\newcommand{\PIc}{\mbox{\textsc{Phase\,Ic}}}
\newcommand{\PII}{\mbox{\textsc{Phase\,II}}}


\newcommand{\fp}{F$_{\rm prompt}$}

\newcommand{\PEC}{PE$_{\rm corr}$}

\newcommand{\order}[1]{\mbox{$\mathcal{O}$(#1)}}

\newcommand{\pic}[5]{
       \begin{figure}[ht]
       \begin{center}
       \includegraphics[width=#2\textwidth, keepaspectratio, #3]{#1}
       \end{center}
       \caption{#5}
       \label{#4}
       \end{figure}
}

\newcommand{\apic}[5]{
       \begin{figure}[H]
       \begin{center}
       \includegraphics[width=#2\textwidth, keepaspectratio, #3]{#1}
       \end{center}
       \caption{#5}
       \label{#4}
       \end{figure}
}

\newcommand{\sapic}[5]{
       \begin{figure}[P]
       \begin{center}
       \includegraphics[width=#2\textwidth, keepaspectratio, #3]{#1}
       \end{center}
       \caption{#5}
       \label{#4}
       \end{figure}
}

\newcommand{\picwrap}[9]{
       \begin{wrapfigure}{#5}{#6}
       \vspace{#7}
       \begin{center}
       \includegraphics[width=#2\textwidth, keepaspectratio, #3]{#1}
       \end{center}
       \caption{#9}
       \label{#4}
       \vspace{#8}
       \end{wrapfigure}
}

\newcommand{\baseT}[2]{\mbox{$#1\times10^{#2}$}}
\newcommand{\baseTsolo}[1]{$10^{#1}$}
\newcommand{\THL}{$T_{\mbox{\tiny 1/2}}$}

\newcommand{\UBI}{$\rm cts/(kg \times yr \times keV)$}

\newcommand{\Uflux}{$\rm m^{-2} s^{-1}$}
\newcommand{\Ucpd}{$\rm cts/(kg \times d)$}
\newcommand{\Uexpo}{$\rm kg\times yr$}

\newcommand{\Qbb}{$Q_{\beta\beta}$}

\newcommand{\validate}{\textcolor{blue}{\textit{(validate!!!)}}}

\newcommand{\improve}{\textcolor{blue}{\textit{(improve!!!)}}}

\newcommand{\missing}{\textcolor{red}{\textbf{...!!!...} }}

\newcommand{\quanta}{\textcolor{red}{\textit{(quantitativ?) }}}

\newcommand{\misscite}{\textcolor{red}{[citation!!!]}}

\newcommand{\missref}{\textcolor{red}{[reference!!!]}\ }

\newcommand{\PC}{$N_{\rm peak}$}
\newcommand{\BIC}{$N_{\rm BI}$}
\newcommand{\PAPR}{$R_{\rm p/>p}$}

\newcommand{\PCR}{$R_{\rm peak}$}


\newcommand{\gline}{$\gamma$-line}
\newcommand{\glines}{$\gamma$-lines}

\newcommand{\gray}{$\gamma$-ray}
\newcommand{\grays}{$\gamma$-rays}

\newcommand{\bray}{$\beta$-ray}
\newcommand{\brays}{$\beta$-rays}

\newcommand{\betas}{$\beta$'s}


\newcommand{\tab}{\textcolor{brown}{Tab.~}}
\newcommand{\eq}{\textcolor{brown}{Eq.~}}
\newcommand{\fig}{\textcolor{brown}{Fig.~}}
\renewcommand{\sec}{\textcolor{brown}{Sec.~}}
\newcommand{\chap}{\textcolor{brown}{Chap.~}}

 \newcommand{\fn}{\iffalse \fi} 
 \newcommand{\tx}{\iffalse \fi} 
 \newcommand{\txe}{\iffalse \fi} 
 \newcommand{\sr}{\iffalse \fi} 

%% file: er_authors.tex
\newcommand{\UofA}{Department of Physics, University of Alberta, Edmonton, Alberta, T6G 2R3, Canada}
\newcommand{\CNL}{Canadian Nuclear Laboratories Ltd, Chalk River, Ontario, K0J 1J0, Canada}
\newcommand{\CIEMAT}{Centro de Investigaciones Energ\'eticas, Medioambientales y Tecnol\'ogicas, Madrid 28040, Spain}
\newcommand{\CU}{Department of Physics, Carleton University, Ottawa, Ontario, K1S 5B6, Canada}
\newcommand{\LNGSA}{INFN Laboratori Nazionali del Gran Sasso, Assergi (AQ) 67100, Italy}
\newcommand{\RHUL}{Royal Holloway University London, Egham Hill, Egham, Surrey TW20 0EX, United Kingdom}
\newcommand{\LU}{Department of Physics and Astronomy, Laurentian University, Sudbury, Ontario, P3E 2C6, Canada}
\newcommand{\UNAM}{Instituto de F\'isica, Universidad Nacional Aut\'onoma de M\'exico, A.\,P.~20-364, M\'exico D.\,F.~01000, Mexico}
\newcommand{\INFN}{INFN Napoli, Napoli 80126, Italy}
\newcommand{\PRISMA}{PRISMA, Cluster of Excellence and Institut f\"ur Kernphysik, Johannes Gutenberg-Universit\"at Mainz, 55128 Mainz, Germany}
\newcommand{\PU}{Physics Department, Princeton University, Princeton, NJ 08544, USA}
\newcommand{\QU}{Department of Physics, Engineering Physics, and Astronomy, Queen's University, Kingston, Ontario, K7L 3N6, Canada}
\newcommand{\RAL}{Rutherford Appleton Laboratory, Harwell Oxford, Didcot OX11 0QX, United Kingdom}
\newcommand{\SL}{SNOLAB, Lively, Ontario, P3Y 1N2, Canada}
\newcommand{\Sussex}{University of Sussex, Sussex House, Brighton, East Sussex BN1 9RH, United Kingdom}
\newcommand{\TRIUMF}{TRIUMF, Vancouver, British Columbia, V6T 2A3, Canada}
\newcommand{\TUM}{Department of Physics, Technische Universit\"at M\"unchen, 80333 Munich, Germany}
\newcommand{\Napoli}{Physics Department, Universit\`a degli Studi ``Federico II'' di Napoli, Napoli 80126, Italy}
\newcommand{\LBLNSD}{Currently: Nuclear Science Division, Lawrence Berkeley National Laboratory, Berkeley, CA 94720}

\affiliation{\UofA}
\affiliation{\CNL}
\affiliation{\CIEMAT}
\affiliation{\CU}
\affiliation{\Napoli}
\affiliation{\LNGSA}
\affiliation{\LU}
\affiliation{\UNAM}
\affiliation{\INFN}
\affiliation{\PRISMA}
\affiliation{\PU}
\affiliation{\QU}
\affiliation{\RHUL}
\affiliation{\RAL}
\affiliation{\SL}
\affiliation{\Sussex}
\affiliation{\TRIUMF}
\affiliation{\TUM}

\author{R.~Ajaj}\affiliation{\CU}
\author{G.\,R.~Araujo}\affiliation{\TUM}
\author{M.~Batygov}\affiliation{\LU}
\author{B.~Beltran}\affiliation{\UofA}
\author{C.\,E.~Bina}\affiliation{\UofA}
\author{M.\,G.~Boulay}\affiliation{\CU}\affiliation{\QU}
\author{B.~Broerman}\affiliation{\QU}
\author{J.\,F.~Bueno}\affiliation{\UofA}
\author{P.\,M.~Burghardt}\affiliation{\TUM}
\author{A.~Butcher}\affiliation{\RHUL}
\author{M.~C\'ardenas-Montes}\affiliation{\CIEMAT}
\author{S.~Cavuoti}\affiliation{\Napoli}\affiliation{\INFN}
\author{M.~Chen}\affiliation{\QU}
\author{Y.~Chen}\affiliation{\UofA}
\author{B.\,T.~Cleveland}\affiliation{\SL}\affiliation{\LU}
\author{K.~Dering}\affiliation{\QU}
\author{F.\,A.~Duncan}\altaffiliation{Deceased.}\affiliation{\SL}
\author{M.~Dunford}\affiliation{\CU}
\author{A.~Erlandson}\affiliation{\CU}\affiliation{\CNL}
\author{N.~Fatemighomi}\affiliation{\SL}\affiliation{\RHUL}
\author{G.~Fiorillo}\affiliation{\Napoli}\affiliation{\INFN}
\author{A.~Flower}\affiliation{\CU}\affiliation{\QU}
\author{R.\,J.~Ford}\affiliation{\SL}\affiliation{\LU}
\author{D.~Gallacher}\affiliation{\CU}
\author{P.~Garc\'ia~Abia}\affiliation{\CIEMAT}
\author{S.~Garg}\affiliation{\CU}
\author{P.~Giampa}\affiliation{\TRIUMF}\affiliation{\QU}
\author{D.~Goeldi}\affiliation{\CU}
\author{V.\,V.~Golovko}\affiliation{\CNL}
\author{P.~Gorel}\affiliation{\SL}\affiliation{\LU}
\author{K.~Graham}\affiliation{\CU}
\author{D.\,R.~Grant}\affiliation{\UofA}
\author{A.\,L.~Hallin}\affiliation{\UofA}
\author{M.~Hamstra}\affiliation{\CU}\affiliation{\QU}
\author{P.\,J.~Harvey}\affiliation{\QU}
\author{C.~Hearns}\affiliation{\QU}
\author{A.~Joy}\affiliation{\UofA}
\author{C.\,J.~Jillings}\affiliation{\SL}\affiliation{\LU}
\author{O.~Kamaev}\affiliation{\CNL}
\author{G.~Kaur}\affiliation{\CU}
\author{A.~Kemp}\affiliation{\RHUL}
\author{I.~Kochanek}\affiliation{\LNGSA}
\author{M.~Ku{\'z}niak}\affiliation{\CU}\affiliation{\QU}
\author{S.~Langrock}\affiliation{\LU}
\author{F.~La~Zia}\affiliation{\RHUL}
\author{B.~Lehnert}\altaffiliation{\LBLNSD}\affiliation{\CU}
\author{X.~Li}\affiliation{\PU}
\author{O.~Litvinov}\affiliation{\TRIUMF}
\author{J.~Lock}\affiliation{\CU}
\author{G.~Longo}\affiliation{\Napoli}\affiliation{\INFN}
\author{P.~Majewski}\affiliation{\RAL}
\author{A.\,B.~McDonald}\affiliation{\QU}
\author{T.~McElroy}\affiliation{\UofA}
\author{T.~McGinn}\altaffiliation{Deceased.}\affiliation{\CU}\affiliation{\QU}
\author{J.\,B.~McLaughlin}\affiliation{\RHUL}\affiliation{\QU}
\author{R.~Mehdiyev}\affiliation{\CU}
\author{C.~Mielnichuk}\affiliation{\UofA}
\author{J.~Monroe}\affiliation{\RHUL}
\author{P.~Nadeau}\affiliation{\CU}
\author{C.~Nantais}\affiliation{\QU}
\author{C.~Ng}\affiliation{\UofA}
\author{A.\,J.~Noble}\affiliation{\QU}
\author{C.~Ouellet}\affiliation{\CU}
\author{P.~Pasuthip}\affiliation{\QU}
\author{S.\,J.\,M.~Peeters}\affiliation{\Sussex}
\author{V.~Pesudo}\affiliation{\CIEMAT}
\author{M.-C.~Piro}\affiliation{\UofA}
\author{T.\,R.~Pollmann}\affiliation{\TUM}
\author{E.\,T.~Rand}\affiliation{\CNL}
\author{C.~Rethmeier}\affiliation{\CU}
\author{F.~Reti\`ere}\affiliation{\TRIUMF}
\author{E.~Sanchez~Garc\'ia}\affiliation{\CIEMAT}
\author{R.~Santorelli}\affiliation{\CIEMAT}
\author{N.~Seeburn}\affiliation{\RHUL}
\author{P.~Skensved}\affiliation{\QU}
\author{B.~Smith}\affiliation{\TRIUMF}
\author{N.\,J.\,T.~Smith}\affiliation{\SL}\affiliation{\LU}
\author{T.~Sonley}\affiliation{\CU}\affiliation{\SL}
\author{R.~Stainforth}\affiliation{\CU}
\author{C.~Stone}\affiliation{\QU}
\author{V.~Strickland}\affiliation{\TRIUMF}\affiliation{\CU}
\author{B.~Sur}\affiliation{\CNL}
\author{E.~V\'azquez-J\'auregui}\affiliation{\UNAM}\affiliation{\LU}
\author{L.~Veloce}\affiliation{\QU}
\author{S.~Viel}\affiliation{\CU}
\author{J.~Walding}\affiliation{\RHUL}
\author{M.~Waqar}\affiliation{\CU}
\author{M.~Ward}\affiliation{\QU}
\author{S.~Westerdale}\affiliation{\CU}
\author{J.~Willis}\affiliation{\UofA}
\author{A.~Zu\~niga-Reyes}\affiliation{\UNAM}

\collaboration{DEAP Collaboration}\email{deap-papers@snolab.ca}\noaffiliation

%% file: intro.tex
Strong evidence suggests dark matter accounts for 84.5\% of the matter
and 26.8\% of the total energy density in the universe \cite{Plank}. Weakly
interacting massive particles (WIMPs) are a favored candidate for
particle-like dark matter and can be searched for with direct
detection via elastic nuclear scattering on target materials in
low-background experiments. DEAP-3600 is searching for WIMP dark
matter with a 3.3~tonne single phase liquid argon (LAr) target, 2070~m
underground (6000 meters water equivalent flat overburden) at SNOLAB in
Sudbury, Canada.  The muon flux at this depth is reduced by about 8 orders of
magnitude compared to sea level~\cite{SNOLAB}.

DEAP-3600 is described in detail in~\cite{DEAPDetector}. It consists
of a spherical acrylic vessel (AV) which holds the target liquid
argon. The spherical volume is filled to approximately 30 cm from the
top. The inside of the acrylic vessel is coated with a 3~$\mu$m-thick
layer of 1,1,4,4-tetraphenyl-1,3-butadiene (TPB) which converts argon
scintillation light to the visible region ($\sim$420 nm).  255 Hamamatsu R5912 high
quantum efficiency photomultiplier tubes (PMTs) view the LAr 
volume via 19-cm diameter light guides (LGs) which are bonded to
the vessel. The space between the light guides is filled with blocks
of alternating layers of high-density polyethylene (HDPE) and Styrofoam. These filler blocks (FBs)
provide neutron shielding and thermal insulation. 
The detector is in a spherical 3.4 m-diameter stainless steel shell which is immersed in
a cylindrical water tank of 7.8~m height and diameter. The data
acquisition, built in the \mbox{MIDAS}~\cite{MIDAS} framework, is triggered and
reads out individual PMT waveforms with single photoelectron
sensitivity. A cross-sectional view of components relevant to this work is given in
\fig~\ref{pic:BackgroundComponents}.

DEAP-3600 detects LAr scintillation light from particle interactions
and separates electronic recoil (ER) backgrounds from nuclear recoil
(NR) WIMP-like events using pulse-shape discrimination
(PSD) \cite{DEAP1PSD}.
In this work, we use the PSD parameter \fp, defined as the ratio of the prompt light to the total light intensity of an event. We use a prompt window 150~ns in length and a total event window of 10~us. 
This window is identical to the analysis in \cite{PRL} and is more appropriate for the high energy events studied in this work than the shorter window recently used in \cite{DEAP2ndWIMP}.
The light intensity is measured in photo electrons (PE) detected across the PMT array. The number of PE in each pulse is determined by dividing the pulse charge by the mean charge of a single PE in the PMT \cite{DEAPPMT}.


NRs dominantly cause the creation of the singlet state of the argon dimer molecule, which
decays in $6-7$~ns. The ER events dominantly yield excited dimers
in the triplet state which decay with a time constant of approximately
1.4~$\mu$s. 
Thus, NR events have an average \fp\ value of approximately 0.7 while ER events have an average \fp\ value of approximately 0.28.

For the purpose of this analysis, ER refers to
events caused by electromagnetic interactions in the argon, including
the passage of fast electrons from $\beta$-decay and from \gray\
interactions such as pair production, Compton scattering, and the
photoelectric effect. Muon tracks in the liquid argon are rare,
result in very large signals and are not considered here. 
Cherenkov radiation can be produced by the same ER interactions relevant for this work 
but is vastly subdominant compared to the scintillation signals at the considered energies, and is thus ignored. 

The argon target used in DEAP is sourced from the earth's atmosphere and contains three stable isotopes: 
\nuc{Ar}{40} (99.60\%), \nuc{Ar}{38} (0.06\%) and \nuc{Ar}{36} (0.33\%), where values in
parenthesis denote the isotopic abundances \cite{nudat2}.  In the even-even nucleus \nuc{Ar}{36} a single
electron capture is energetically forbidden but a decay via double
electron capture - a second-order process similar to double
$\beta$-decay - can occur. Currently, the best limit on the half-life for the
neutrinoless decay mode in \nuc{Ar}{36} is \THL \baseT{>3.6}{21}~yr (90\% credibility interval)
\cite{Gerda16}.  

The argon target also contains three unstable but long-lived cosmogenic isotopes: 
\nuc{Ar}{42} (\THL=$32.9\pm1.1$~yr), \nuc{Ar}{39}
(\THL=$269\pm3$ yr) and \nuc{Ar}{37} (\THL=$35.04\pm0.4$ d).  The vast
majority of events in DEAP-3600 are ERs from the $\beta$-decay
of \nuc{Ar}{39}, intrinsic to atmospheric argon with a measured specific activity
of $1.01 \pm 0.02\mbox{(stat)} \pm 0.08\mbox{(syst)}$~Bq/kg (WARP collaboration \cite{Ar39_warp})
and $0.95\pm0.05$~Bq/kg (ArDM collaboration \cite{Ar39_ArDM}).
 \nuc{Ar}{39} dominates the ER
background by approximately a factor of
100 up until its endpoint at $565\pm5$~keV. 
The $\beta$-decay of \nuc{Ar}{42} and its daughter \nuc{K}{42}
have approximately 4 orders of magnitude less specific activity than
\nuc{Ar}{39} and its precise concentration in atmospheric argon is
not well determined \cite{Ashitkov98,Ashitkov03,Ar42,Barabash16}. 
Especially the daughter decay of \nuc{K}{42} is a dominant background in neutrinoless double beta decay experiments using argon as shielding such as GERDA and LEGEND.
A dedicated analysis of the \nuc{Ar}{42} activity in atmospheric argon is presented in this work. 
The electron capture
of \nuc{Ar}{37} has a relatively short half-life compared to
experimental time scales and only emits X-rays up to 2.8~keV. It is
not considered in this work.

The main direct backgrounds for the WIMP search are (1) $\alpha$-decays
on the surface of the detector in which only a fraction of their
energy is deposited in scintillating material, (2) 
neutron interactions, which produce recoiling argon nuclei similar to WIMP interactions,
and (3) ER events that are misidentified as NR. DEAP-3600 was designed
so that each of these background components accounts for less than 0.2
events in a 3000~\Uexpo\ exposure after fiducialization for the WIMP
search \cite{DEAPDetector}.

The dominant source of neutron emission in the detector is from 
 ($\alpha$,n) reactions in the borosilicate glass of the
PMTs. The activity from $\alpha$-decays in the \nuc{U}{238} and \nuc{Th}{232} chains
can be constrained by measuring the \grays\ within the ER background
model. In addition, the overall neutron flux in the surrounding
detector material results in (n,$\gamma$) reactions producing \grays\ with energies up to approximately 10~MeV, which are measured. 

In this paper a model is developed for the ER backgrounds in
DEAP-3600 for data collected in the first year of operation. 
Expected background contributions are fit to
the data in a Bayesian framework, in order to obtain information on the
activity or specific activity of detector components. In \sec
\ref{sec:dataselection} the dataset and cuts for pile-up rejection and data cleaning are explained,
along with the energy calibration of ER events.  
\sec \ref{sec:backgroundsources} discusses in detail the 
components used in the model and how they were simulated in
a Geant4-based \cite{Geant4} Monte Carlo (MC) simulation framework. The analysis of the background
decomposition is described in \sec \ref{sec:backgrounddecomposition}. 
The results and the validation of the analysis are given in \sec \ref{sec:results} and \ref {sec:externalgammaverification}.
Finally, \sec \ref{sec:Ar42K42} discusses a
dedicated analysis of the \nuc{Ar}{42} concentration in atmospheric
argon, looking at the contribution of its daughter \nuc{K}{42} to the
ER background above the 2614.5~keV \gline\ of \nuc{Tl}{208}.

%% file: datasetcuts.tex
This paper uses an open data set taken between November 1, 2016 and
October 31, 2017. The lifetime for physics data taken in that timeframe is 247.2 days after data quality checks\footnote{This paper includes approximately 0.5 days more exposure than \cite{DEAP2ndWIMP}. This corresponds to short runs in which the trigger efficiency at low energy was not sufficiently well determined for WIMP analysis.}.


This analysis uses a small set of cuts designed to remove instrumental
artifacts and event pile-up, while keeping scintillation events with high acceptance and
little energy- and position-dependence within the detector. 
Therefore, fiducial volume cuts are not used.

Events are selected under the following conditions. 
(1) The trigger is derived from LAr PMTs, excluding muon veto triggers and pulser events.
(2) A series of low-level criteria are met: 
no pile-up between different trigger types,
the DAQ was in a ready state,
the pulse-finding algorithm was successful, 
and events had a stable baseline. 
The low-level cuts remove $< 0.1$\% of events. 
(3) No pile-up was detected by any of the following pile-up criteria: 
a template fit finds more than one scintillation-like signal in the recorded event,
the reconstructed ``time zero'' of an event lies outside a narrow window defined by the trigger time,
the recorded PMT signals before the trigger time have $\ge 3$~PE, 
and the event occurs less than 20~$\mu$s after the previous event.
The pileup cuts reliably remove pile-up events separated by more than 0.5~$\mu$s. 
With a trigger rate of 3170~Hz, the detected and removed pile-up fraction is 5.4\%. The fraction of un-detected pile-up within 0.5~$\mu$s is 0.15\% and thus negligible. 
The removed pile-up fraction is an input to the detection efficiency in the background model on which we assume a conservative 10\% systematic uncertainty.

\begin{figure}
  \centering 
  \includegraphics[width=\columnwidth]{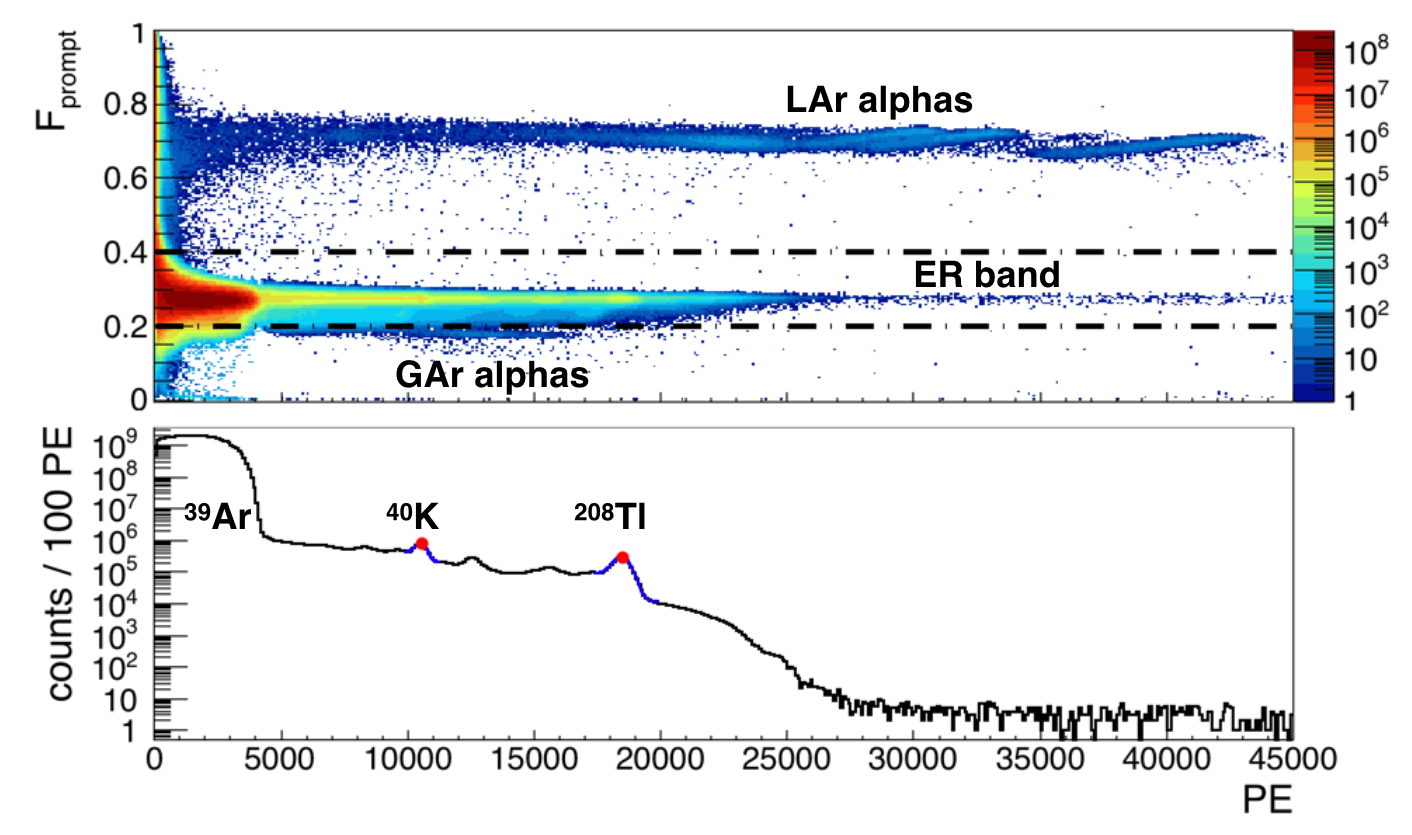} 
  \caption{\label{pic:EnergyCalibration1}
  Top: \fp\ vs.\ PE spectrum of the dataset after sequence of cuts and before
  energy calibration. The horizontal lines illustrate the
  selected \fp\ region for this work. The color scale denotes counts per 100~PE and 0.005~\fp\ bin.
  Bottom: Selected ER events. The
  fit functions of the two prominent background \glines\ at 1460.8~keV
  from \nuc{K}{40} and at 2614.5~keV from \nuc{Tl}{208} are shown as
  well as the mean value.}
\end{figure}

All selected events in the dataset are shown in a
\fp\ vs.\ PE plot in the top panel of \fig \ref{pic:EnergyCalibration1}.
A NR band can be seen at high \fp\ values between 0.6 and 0.8, mainly
populated by $\alpha$-decays. An ER band emerges at low \fp\ values
between 0.2 and 0.4. Below the ER band 
is a population of events caused by $\alpha$-decays in the gaseous argon above the liquid level, most prominent between approximately 10000 to 15000~PE.
The horizontal dashed lines illustrate the ER event selection for
this work. The lower panel of \fig \ref{pic:EnergyCalibration1} shows
the projection of the selected ER events with the 1460.8~keV \gline\
from
\nuc{K}{40} and the 2614.5~keV \gline\ from \nuc{Tl}{208}
highlighted.

%% file: calibration.tex
The DAQ system, including PMT gain and amplification, is optimized for
small signals from low-energy WIMP interactions. Thus the energy
calibration used for the WIMP analysis is not necessarily appropriate 
for high-energy signals as clipping of the waveform readout and PMT
saturation must be considered.

\begin{figure}[h]
        \centering
        \includegraphics[width=\columnwidth]{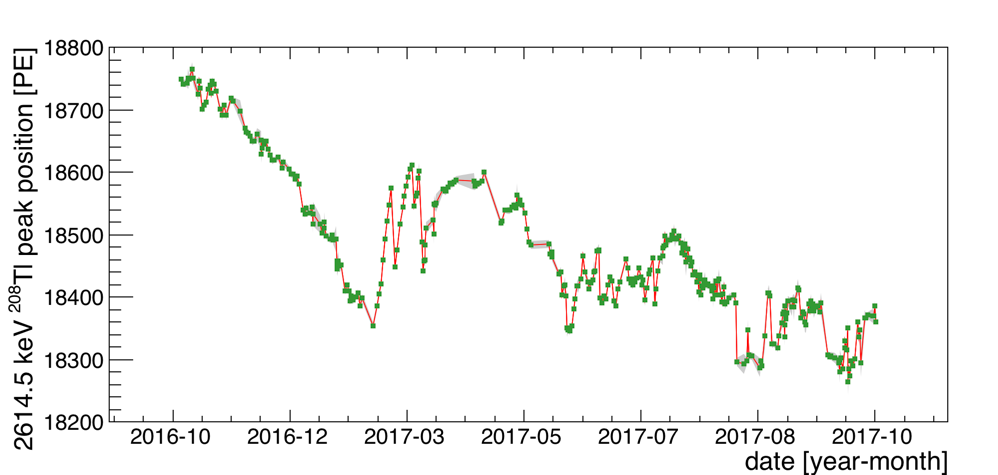}
         \caption{\label{pic:pdf_g_peak1_pos} The position of the 2614.5~keV peak from \nuc{Tl}{208} 
           for runs in the dataset. 52 short runs with
           insufficient statistics out of 402 total runs are not
           shown. Uncertainties of the peak position are shown
           with a grey error band in the order of 20~PE or 0.1\%.. }
\end{figure}

\begin{figure}[h]
        \centering
                \includegraphics[width=\columnwidth]{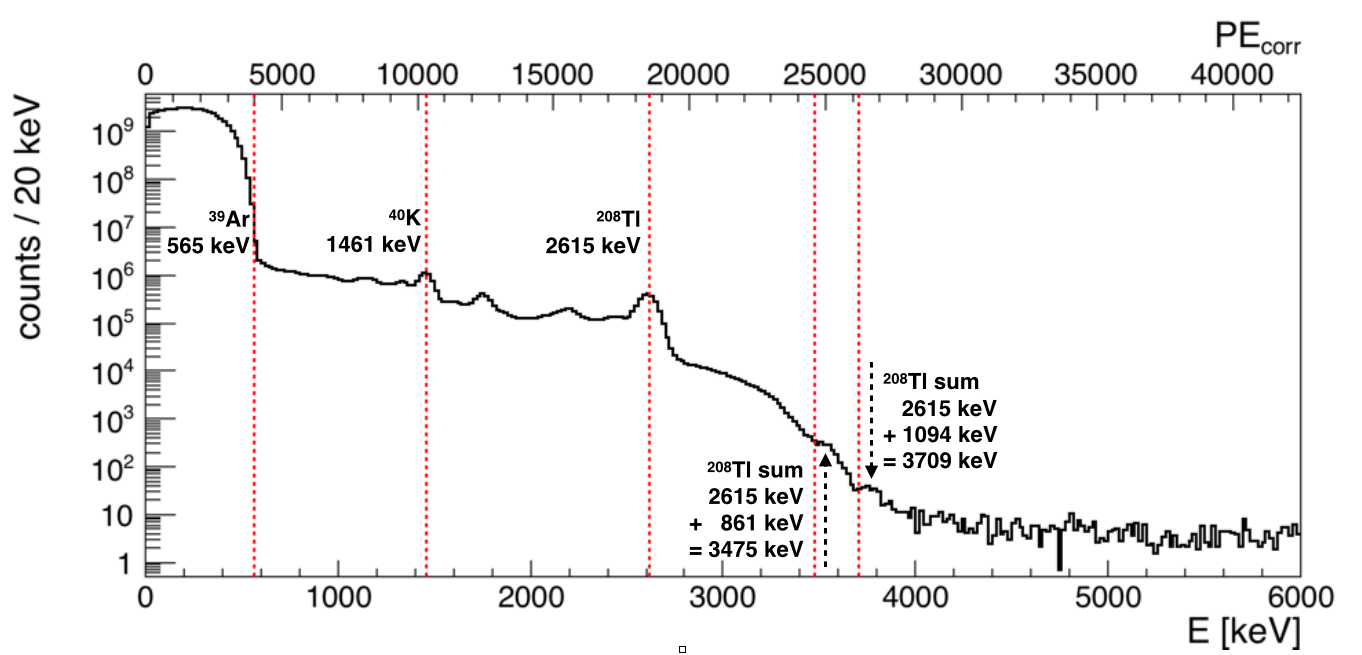}
         \caption{\label{pic:pdf_keVandPECorr} Energy spectrum
           calibrated in keV and $\rm PE_{corr}$. Prominent energies
           are shown as vertical dashed lines. The summation peaks of
           \nuc{Tl}{208} are clearly visible and indicated with
           arrows. In the energy calibration based on single
           \gray\ peaks, the summation peaks are mis-calibrated as
           discussed in the text.  }
\end{figure}

In this work an effective calibration is used. The \glines\ from
\nuc{K}{40} and \nuc{Tl}{208} are used to define a quadratic function
\begin{equation}
E = 0 + p_1 \times {\rm PE} + p_2 \times {\rm PE}^2\ . \label{eq:1}
\end{equation}
Other prominent \glines, e.g.\ from \nuc{Bi}{214}, have secondary \glines\ within the
peak resolution and are not suitable for precise calibration.  
In each run, the
\glines\ are fitted with a Gaussian peak shape and an
empirical background function consisting of two constants connected with an
error function fixed to the mean and the width of the
peak shape. The means of both \glines\ are extracted for each run with
sufficient statistics. The mean peak position of the 2614.5~keV
\gline\ is plotted for 350 out of 402 runs in \fig \ref{pic:pdf_g_peak1_pos}, illustrating energy scale variations up to
2.5\%. 
The effects of these variations are corrected by calibrating the energy scale on a run-to-run basis. 
For 52 runs with insufficient statistics, the previous run calibration is used. 
The spectrum obtained after calibrating each run to keV and then summing up the individual spectra is shown in
\fig \ref{pic:pdf_keVandPECorr}. 
For use in the analysis, the calibrated energy spectrum is converted back into corrected photoelectrons (PE$_{\rm corr}$) in which the quadratic term in \eq \ref{eq:1} is removed. The PE$_{\rm corr}$ energy scale, on which the fit is later performed, is also shown in \fig \ref{pic:pdf_keVandPECorr}.

Monte Carlo simulations are used to determine the energy depositions in the LAr for each background component in the model.
This energy deposition spectrum is convolved with an analytic detector
response function (for energy scale and resolution) to obtain a spectrum that is compared with the linearized data in PE$_{\rm corr}$.

The energy calibration described above is based on peaks from single \grays. 
Events with more complicated topologies (e.g.\ summation peaks) have a
more diffuse distribution of light and are thus less affected by
non-linearities. Such an effect is observed for sources of
\nuc{Tl}{208} close to the LAr in which summations of the 2614.5~keV \gray\ with \grays\ at 583.2~keV,
860.6~keV or 1093.9~keV can occur as shown in \fig \ref{pic:pdf_keVandPECorr}. 

\begin{figure}
\centering 
\includegraphics[width=\columnwidth]{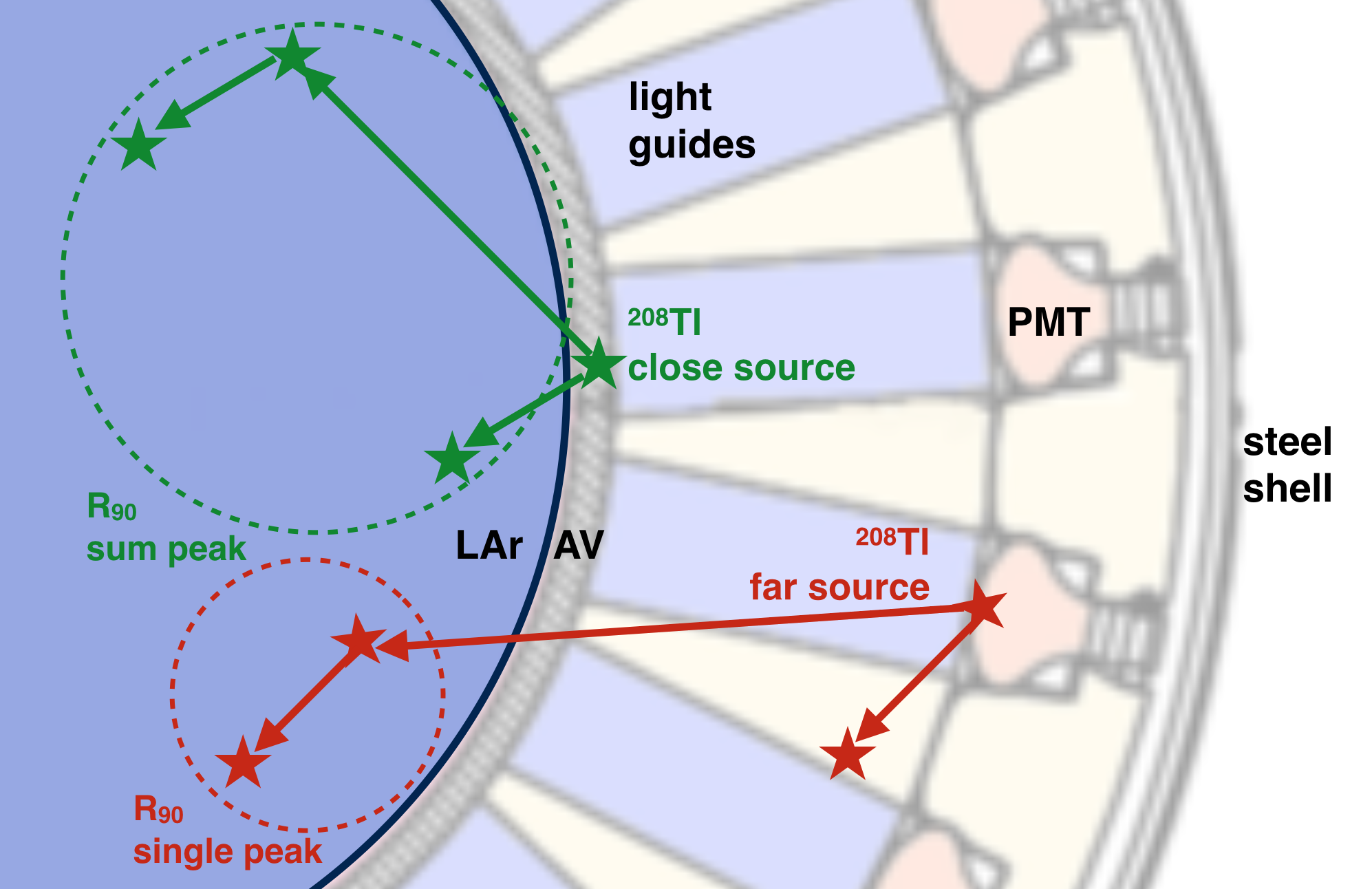}
\caption{\label{pic:SaturationIllustration} Illustration of different saturation effects for single \grays\ and summation \gray\ interactions. For close sources of \nuc{Tl}{208}
(green) it is more likely that two \grays\ enter the LAr creating a
broader spacial distribution of energy depositions and scintillation light than
for single \grays\ (red). A more diffuse light distribution will expose
the PMT array more evenly and results in overall less
saturation. These topologies are distinguished with the size of the
$R_{90}$ parameter as illustrated by the circles.}
\end{figure}

\begin{figure}[h]
  \centering
     \includegraphics[width=\columnwidth]{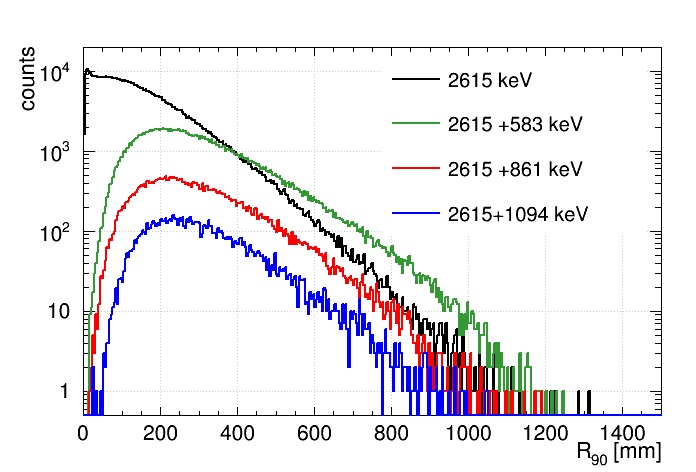}
     \includegraphics[width=\columnwidth]{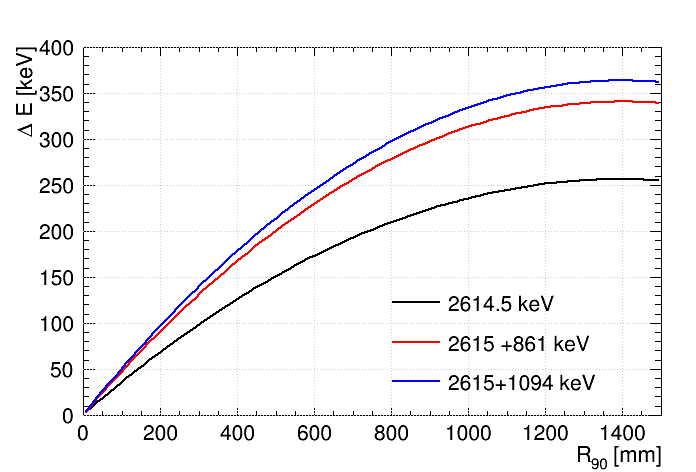}              
     \caption{\label{fig:topologycorrection} Top: $R_{90}$ distribution at single and summation peak energies for \nuc{Tl}{208}\ decay from Monte Carlo simulations. Bottom: The average shift in measured energy caused by the topology. }
\end{figure}

The effect of topology on the measured energy is empirically modeled in Monte Carlo with a
correction term based on the radius of a sphere containing 90\% of the
deposited energy ($R_{90}$) as illustrated in \fig
\ref{pic:SaturationIllustration}. The $R_{90}$ distributions for
simulated events from \nuc{Tl}{208} decays in the AV bulk are shown
in \fig \ref{fig:topologycorrection} (top). The single \gline\ at 2614.5~keV shows significantly lower $R_{90}$ values peaking at around 10~mm, with an exponential tail towards larger spheres. The summation peaks, on the other hand, show wider distributions peaking at around 200~mm which are more dependent on the detector geometry and the distance of the source. This behavior is used to construct an event by event energy correction $E' = E + \Delta E$  in which $\Delta E$
is based on an empirical function of $R_{90}$ and energy:
\begin{equation}
 \Delta E(E,R_{90}) = E \times (q_0 + q_1 \times R_{90} + q_2 \times R_{90}^2 )\ . \label{eq:2}
\end{equation}
The parameters $q_0$, $q_1$, and $q_2$ were manually determined to best match an example spectrum from a \nuc{Tl}{208} source close to the LAr to data and set to 0, \baseT{1.4}{-4} and \baseT{-5.0}{-8}, respectively. The energy shift for \nuc{Tl}{208} single and
sum peaks is shown in \fig \ref{fig:topologycorrection} (bottom).
A Gaussian resolution ($\sqrt{r \times \rm PE_{corr}}$) is applied to the simulated spectra after the energy correction which already introduces intrinsic smearing for different event topologies. The effective parameter $r$ is later determined in an energy response fit. 

Using an analytical effective detector response allows for (1) high
statistics Monte Carlo simulations without the computational cost of photon
tracking; (2) simultaneous fitting of background components and energy
scale/resolution; and (3) different energy scales to be applied to
internal and external components to model position dependencies and
non-linearities at higher energies. The mathematical construction of the
empirical model conserves the total number of events for the activity
estimation. However, the correction of the energy scale as well as the linearization of the response results in {\it effective parameters} which should only be interpreted in achieving an optimal match of simulations to data.

%% file: backgroundsources.tex
The vast majority of ER background in DEAP-3600 is due
to $\beta$-electron or \gray\ interactions with the LAr produced by nuclear
decays. 
The muon background is subdominant at the considered energies due to the low muon rate at SNOLAB and the low probability of short track length in the monolithic spherical detector.
Background components can
be roughly divided into sources internal and external to the LAr. 
Internal backgrounds have nearly 100\% probability of generating a
signal, whereas external sources have a much lower
probability.
For internal sources, the large LAr volume functions as a calorimeter,
mostly detecting the total \gray\ and $\beta$ energy of a decay. 
For external sources, primarily \grays\ are detected. However, for sources
close to the LAr such as the AV itself, partial $\beta$ and
bremsstrahlung components can contribute to the observed energy. In
addition, a small amount of Cherenkov light can add to the LAr
scintillation light for simultaneous energy depositions in LAr and non-scintillating optically transparent media such as the AV or LG acrylic
as well as the PMT glass.  

Different source positions of the same nuclide can 
produce degenerate experimental signatures in DEAP-3600. For instance,
the single 1460.8~keV \gray\ from the electron capture (EC) branch of \nuc{K}{40}
does not allow disentangling the activity of multiple components at
different radii. 
In contrast, the multi-$\gamma$ cascade of
\nuc{Tl}{208} decays allows for the determination of the distance between the
source and the argon, using different probabilities of summation \glines\ 
and different peak-to-peak ratios due to more strongly
attenuated low energy \grays. 
Contributions from the \nuc{U}{235} decay chain are generally omitted in this analysis since its natural abundance is about a factor of 20 below that of \nuc{U}{238} and the chain does not contain \grays\ with more than 5\% emission probability above the \nuc{Ar}{39} $\beta$ endpoint.

\begin{figure}[h]
        \centering
                \includegraphics[width=\columnwidth]{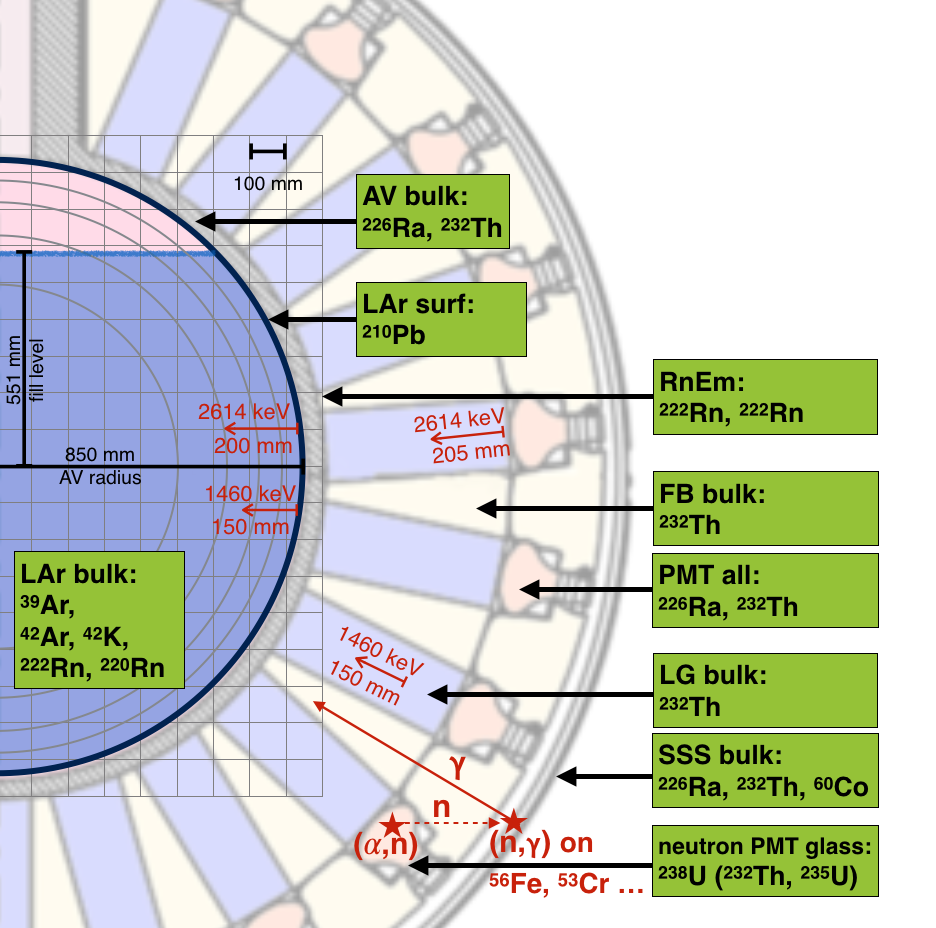}
         \caption{\label{pic:BackgroundComponents} 
         Illustration of considered background components in the detector setup (to scale). 
         Shown are background source positions from  internal to far sources: liquid argon bulk (LAr bulk), TPB-acrylic vessel interface (LAr surf), acrylic vessel bulk (AV bulk), radon emanation towards the outer AV surface (RnEm) light guide bulk (LG bulk), filler block bulk (FB bulk), PMT components (PMT all), borosilicate glass of PMTs (PMT glass), stainless steel sphere bulk (SSS bulk).
         Also illustrated is the fill level, the attenuation length of 2614.5~keV and 1460~keV \grays\ in acrylic and LAr as well as processes of neutron captures resulting in \grays. 
         }
\end{figure}

The detector materials and their dominant background contributions are 
illustrated in \fig \ref{pic:BackgroundComponents}. The material assay of individual components \cite{GammaSpectroscopyAssay} is discussed in the appendix and listed in \tab \ref{tab:ScreeningResults1} and \ref{tab:PMTScreening}.

\subsection{Internal Sources}

Internal sources are expected to be homogeneously distributed in the bulk of the LAr. 

{\bf \nucbf{Ar}{39} LAr bulk and \nucbf{Ar}{42}/\nucbf{K}{42} LAr bulk}: 
\nuc{Ar}{39} and \nuc{Ar}{42} are long lived radioactive isotopes of argon, which have an approximately constant rate over the timeframe of the dataset.
They are produced cosmogenically and thus their specific activities are
expected to be equal in any argon batch extracted from the
atmosphere. Both decays are unique first-forbidden $\beta$-decays to the
ground state of the daughter with Q-values of \unit[$565\pm5$]{keV}
and \unit[$599\pm6$]{keV}, respectively. \nuc{Ar}{39} has a four order
of magnitude higher specific activity than \nuc{Ar}{42} and dominates
the low energy spectrum in DEAP-3600. However, \nuc{Ar}{42} decays are
followed by decays of \nuc{K}{42} (\THL=\unit[$12.355\pm0.007$]{h}), with a
Q-value of \unit[$3525.2 \pm0.2$]{keV}, which dominates the energy
spectrum above 2.6~MeV.

\nuc{Kr}{85} (\THL=\unit[$10.739\pm0.014$]{yr}), a $\beta$ emitter with a
Q-value of 687.0~keV was previously found in LAr
detectors \cite{Ar39_warp,DS_532d}; however, its specific activity
depends on the LAr purification and is thus unique to each experiment or measurement. 
A dedicated analysis has been performed searching for the \nuc{K}{85} decay to the
514.0~keV excited state of \nuc{Rb}{85}, with a probability of 0.434\%
and a half-life of 1.02~$\mu$s. This decay mode would result in two
correlated scintillation peaks in the same event trace. No evidence
of \nuc{Kr}{85} has been observed in the data, and an upper limit of
1.5~mBq/kg (90\% CL) has been set. \nuc{Kr}{85} is thus neglected in the
ER model as subdominant compared to \nuc{Ar}{39} and other
signals.  

{\bf \nucbf{Rn}{220} LAr bulk and \nucbf{Rn}{222} LAr
bulk}: \nuc{Rn}{220} (\THL=\unit[$55.6 \pm0.1$]{s})
and \nuc{Rn}{222} (\THL=\unit[$3.8235\pm0.0003 $]{d}) enter the LAr via the
piping of the process system and enhance the observed activity in the lower parts of
the \nuc{U}{238} and \nuc{Th}{232} chains, respectively. The activities
of these isotopes are obtained from a dedicated {\it in-situ} $\alpha$ background
analysis \cite{PRL}. In this dataset the process systems were isolated
from the detector and the observed $\alpha$-decays show a constant rate, 
indicating a source in the piping after the last valve or in the detector neck. 

The ER background response is mainly due to $\beta$-decay daughters of 
\nuc{Bi}{214} and \nuc{Tl}{208},
which have high Q-values of \unit[3270]{keV} and \unit[4999]{keV},
respectively. The \nuc{Rn}{222} chain is broken at \nuc{Pb}{210}  (\THL=\unit[$22.2\pm0.2 $]{yr}) which
does not reach equilibrium within the dataset. The chain
below \nuc{Pb}{210} is not considered in the background model.

\subsection{LAr Surface/TPB Layer/Acrylic Surface} 

The analysis in \cite{PRL} identifies \nuc{Po}{210} decays, fed by \nuc{Pb}{210}, on the TPB-acrylic interface and
potentially in the first \unit[80]{$\mu$m} of the acrylic bulk. For
the ER background model considering $\beta$ and \gray\ emission, these
scenarios are approximately identical. 

{\bf \nucbf{Pb}{210} LAr surf}: Only the \nuc{Pb}{210}
daughter \nuc{Bi}{210}, with a Q-value of \unit[1162]{keV}, is
considered in the model. 2.0~mBq originate from the TPB-acrylic
interface and $<3.3$~mBq from the acrylic bulk which are combined into one component.

\subsection{External Sources}

\subsubsection{AV Bulk}

{\bf \nucbf{Ra}{226} AV bulk and \nucbf{Th}{232} AV bulk}: All $\beta$ and
\gray\ emitting isotopes below \nuc{Ra}{226} in the \nuc{U}{238} decay
chain are considered as well as all such isotopes in the \nuc{Th}{232}
chain, since even low energy \grays\ can enter the LAr from the AV.  No
evidence of \nuc{C}{14} has been found in the acrylic using accelerator mass spectrometry \cite{AMSOttawaU} and \nuc{C}{14} is
subsequently not considered in the model.
 
{\bf \nucbf{Rn}{220} RnEm and \nucbf{Rn}{222} RnEm}: Radon can, in
principle, emanate from materials inside the stainless steel shell and
freeze out at the cold outer AV surface. This is a hypothetical
component which is not included in the model but investigated as a systematic uncertainty.

\subsubsection{LG Bulk and FB Bulk}

{\bf \nucbf{Th}{232} LG bulk and \nucbf{Th}{232} FB bulk}: Both the
LGs and FBs are known from screening to be low in radioactive contaminants. 
However, summation \glines\ from \nuc{Tl}{208} can have a
dominant effect in the spectrum above 2614.5~keV. The slightly
different geometry of the LGs and FBs allows for some breaking of the
degeneracy between them.
Isotopes with a small Q-value and consequently low \gray\ energies such as \nuc{Pb}{210} in
the \nuc{Ra}{226} chain and \nuc{Ra}{228} in the \nuc{Th}{232} chain
cannot practically contribute to the spectrum in the detector and are
not considered.

\subsubsection{PMTs} 

The PMTs have the highest specific activity in DEAP-3600. PMT glass,
PMT inner components, and PMT mounting components are individually
tracked but summed and
treated as one component in the fit, denoted ``PMT all''. 
The individual PMT components are listed in \tab \ref{tab:PMTScreening} in the appendix.  

{\bf \nucbf{Ra}{226} PMT all, \nucbf{Th}{232} PMT all, and \nucbf{K}{40}
PMT all}: Activities from the primordial decay chains and \nuc{K}{40}
are considered.

{\bf Neutron PMT glass}: Neutrons created through ($\alpha$,n) reactions
in the borosilicate PMT glass are captured in the surrounding material
and create a variety of \grays\ in a variety of locations through
(n,$\gamma$) reactions. This process is modeled in the simulation by generating
neutrons calculated for \nuc{U}{238} in borosilicate glass with SOURCES 4C \cite{SOURCES4C}. 
The resulting ER spectrum from \grays\ from neutron capture processes is mostly flat, spanning an energy
range up to 10~MeV. The only distinct feature is the
2224.5~keV \gline\ from captures in \nuc{H}{1}. 
($\alpha$,n) reactions from \nuc{U}{238} are the dominant neutron source in DEAP-3600 and the resulting \gray\ spectrum serves as an approximated template in the model. 
Simulations indicate that spectra for ($\alpha$,n) and spontaneous fission neutrons from the \nuc{U}{238}, \nuc{U}{235}, and \nuc{Th}{232} decay chains are similar to this template. In this analysis it serves only as a high-energy \gray\ spectrum and is not used for a neutron background prediction in DEAP-3600.


\subsubsection{The Stainless Steel Shell bulk} 
Of the detector components considered in the model, the SSS is the furthest away from the active volume, but it also has the largest total mass. At the distance of the
SSS, only isotopes with sufficiently high energy \grays\
create a detector response.

{\bf \nucbf{Ra}{226} SSS bulk and \nucbf{Th}{232} SSS bulk}: For
the \nuc{Ra}{226} chain, only \nuc{Bi}{214} is simulated; for
the \nuc{Th}{232} chain, \nuc{Ac}{228} and \nuc{Tl}{208} are
simulated.

{\bf \nucbf{Co}{60} SSS bulk}: 
In addition to the primordial isotopes, significant amounts of \nuc{Co}{60} are present in the SSS and \nuc{Co}{60} is therefore considered in the model.



\subsubsection{Summary}

The components entering the background model are summarized
in \tab \ref{tab:InputActivities}. 
The first column shows the background component indicating the radioactive isotope or decay chain in a certain detector part. Only overall dominant components are listed and their use in the model is indicated in the second column.
For each component, only dominant isotopes are simulated which are shown in the third column.
The fourth column shows the total expected activity in the DEAP-3600
materials and is used as prior information in the fit. 
The quoted uncertainties combine the uncertainty on the assayed specific activity with the uncertainty on the total mass of the component.
See also \cite{DEAPDetector} for more information.

\begin{table*}
\begin{center}
\caption{Screening results for ER background components used in this document. The columns from left to right denote: the head isotope of the decay chain and the location of the component; the use of the component in the model; the simulated isotopes of the chain; the prior knowledge of the total activity in DEAP-3600; and a reference. The use of each component is either (F) free parameter (C) set to constant (N) not included or (D) used in a dedicated analysis.  The line ``neutron PMT glass'' refers to n-capture \grays\ from all detector materials resulting from ($\alpha$,n) reactions in the PMT glass from all $\alpha$'s in the \nuc{U}{238} chain.}
\label{tab:InputActivities}
\begin{ruledtabular}
\begin{tabular}{lcllc} 
Component & Included & Simulated & Total activity & Reference \\ 
      & in model? & isotopes & [Bq] &  \\ \colrule
\nuc{Ar}{39} LAr bulk & F & \nuc{Ar}{39} & $3282\pm340$
      & \cite{Ar39_warp} \\
\nuc{Ar}{42}/\nuc{K}{42} LAr bulk & F
      & \nuc{Ar}{42} , \nuc{K}{42} & --- & --- \\ 
\nuc{Rn}{222} LAr bulk & C & \nuc{Pb}{214}, \nuc{Bi}{214}
      & \baseT{(5.9\pm0.7)}{-4} & \cite{PRL} \\ 
\nuc{Rn}{220} LAr bulk & F & \nuc{Pb}{212}, \nuc{Bi}{212}, \nuc{Tl}{208}
      & \baseT{(8.5\pm4.9)}{-6} & \cite{PRL} \\ 
\colrule 
\nuc{Pb}{210} LAr surf & C & \nuc{Pb}{210}, \nuc{Bi}{210}
      & \baseT{(2.2\pm0.4)}{-3} & \cite{PRL} \\ 
\colrule 
\nuc{Ra}{226} AV bulk & F & \nuc{Pb}{214}, \nuc{Bi}{214}, \nuc{Pb}{210}, \nuc{Bi}{210} &
      $<0.08$ & [screening] \\ 
\nuc{Th}{232} AV bulk & F & \nuc{Ra}{228}, \nuc{Ac}{228}, \nuc{Pb}{212}, \nuc{Bi}{212}, \nuc{Tl}{208}
      & $<0.22$ & [screening] \\ 
\nuc{K}{40} AV bulk & N & \nuc{K}{40} & $<2.5$ & [screening] \\ \colrule 
\nuc{Rn}{222} RnEm & D
      & \nuc{Bi}{214} & $<1$& \cite{DEAPDetector}\ \\ 
\nuc{Rn}{220} RnEm & D & \nuc{Tl}{208} & $<1$ & \cite{DEAPDetector} \\ 
\colrule 
\nuc{Ra}{226} LG bulk & N & \nuc{Pb}{214}, \nuc{Bi}{214}, \nuc{Bi}{210} & $<0.4$ &
      [screening]\\ 
\nuc{Th}{232} LG bulk & F & \nuc{Ac}{228}, \nuc{Pb}{212}, \nuc{Bi}{212}, \nuc{Tl}{208} &
      $<1.3$ & [screening] \\ 
\nuc{K}{40} LG bulk & N & \nuc{K}{40} & $<4.6$ & [screening] \\ \colrule 
\nuc{Ra}{226} FB bulk & N
      & \nuc{Pb}{214}, \nuc{Bi}{214}, \nuc{Bi}{210} & $<1.5$ &
      [screening] \\ 
\nuc{Th}{232} FB bulk & F
      & \nuc{Ac}{228}, \nuc{Pb}{212}, \nuc{Bi}{212}, \nuc{Tl}{208} &
      $<0.9$ & [screening]\\ 
\nuc{K}{40} FB bulk & N & \nuc{K}{40} &
      $<9.6$ &[screening] \\
\colrule
\nuc{Ra}{226} PMT all & F     &   \nuc{Pb}{214}, \nuc{Bi}{214}, \nuc{Bi}{210} 
              &   $216 \pm 24$ & [screening]   \\
\nuc{Th}{232} PMT all & F     &   \nuc{Ac}{228}, \nuc{Pb}{212}, \nuc{Bi}{212}, \nuc{Tl}{208}      
              &  $39 \pm 4$     & [screening]  \\
\nuc{K}{40} PMT all   & F    &    \nuc{K}{40}  &  $454 \pm 33$  & [screening]   \\
neutron PMT glass   & F &      See caption                 & ---  & ---     \\
\colrule
\nuc{Ra}{226} SSS bulk & F       &   \nuc{Bi}{214}                & $10.6 \pm 5.8 $ 
              &  [screening] \\
\nuc{Th}{232} SSS bulk & F      &   \nuc{Ac}{228}, \nuc{Tl}{208}    & $9.7 \pm 5.6 $    
              & [screening] \\
\nuc{Co}{60} SSS bulk  & F      &   \nuc{Co}{60}                 & $78 \pm 11 $   
             & [screening] \\
\end{tabular} 
\end{ruledtabular}
\end{center}
\end{table*} 

\subsection{Monte Carlo Simulations}

The DEAP-3600 geometry, including the surrounding hall at SNOLAB, is
implemented in the simulation and analysis framework RAT \cite{RAT} which is based on Geant4 \cite{Geant4}
version 9.6.p02. 
The full optical model developed for the dark-matter
search analysis is not used here. Radioactive decays are modeled
with the event generator Decay0~\cite{Decay0} for each background component listed
in \tab \ref{tab:InputActivities}.  The number of generated decays
varies depending on the isotope and its distance from the LAr target.
The simulated number of decays may be more or less than the expected number of
decays in the data set. For example, for \nuc{Ar}{39} and \nuc{K}{42} in the
LAr bulk and \nuc{Bi}{214} and \nuc{Tl}{208} in the PMTs,
about \baseT{7}{10}, \baseT{6}{6}, \baseT{5}{9}, and \baseT{8}{8}
decays are expected in the dataset, respectively, whereas a total
of \baseT{1}{7}, \baseT{1}{7}, \baseT{4.1}{7}, and \baseT{4.3}{7}
decays are simulated for these components.  Geant4 hit positions and
energy depositions are stored in optically sensitive volumes for
further post-processing, including the empirical correction of the energy
scale based on the event topology. 
Isotopes in the same decay chain are summed together, accounting for their branching ratios when secular equilibrium is assumed: e.g.\ the simulation of \nuc{Tl}{208} is added with a
weight of 0.3594 to the \nuc{Th}{232} chain. 
The Monte Carlo energy depositions in the LAr are shown in \fig \ref{pic:pdf_cpdBeforeCorrection_test} for selected background components. These are the base of the ER model and show the underlying spectral features which are ultimately smeared out by the detector response. The vertical scaling illustrates the probability of a detected signal for a given decay of the head isotope in the background component.

%% file: decomposition.tex
Spectra from MC simulations of the individual background components in \tab \ref{tab:InputActivities} are matched to the PE$_{\rm corr}$
spectrum in the ER band in a combined fit. 
All components are scaled to livetime with the activity as a fixed or floated parameter in the fit. A floated activity can be constrained or
free, depending on the prior information.

\begin{figure*}[h]
        \centering
                \includegraphics[width=0.95\textwidth]{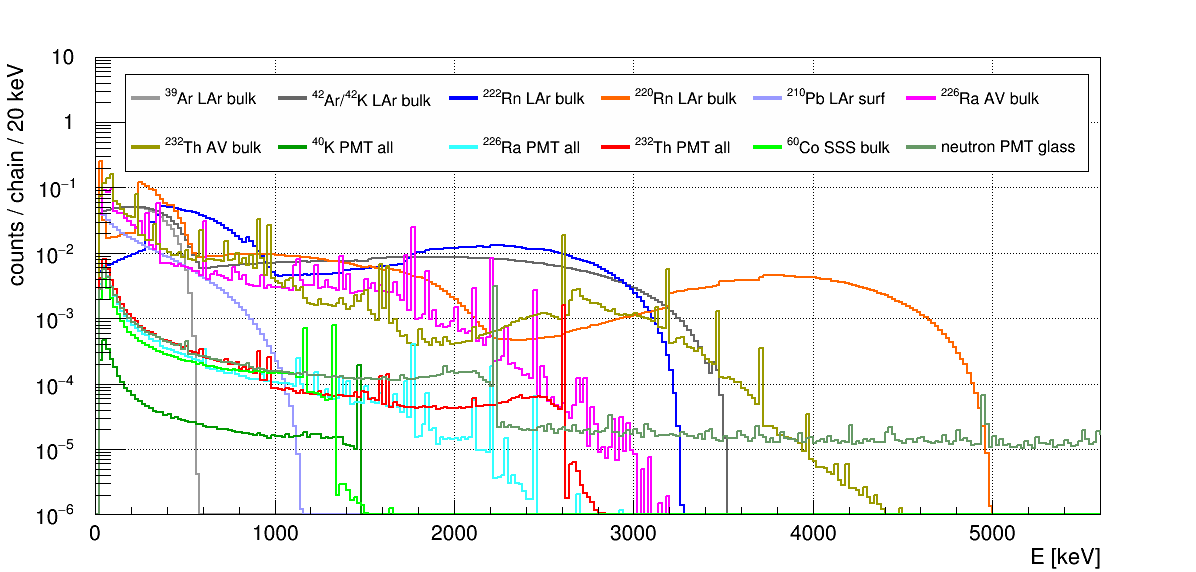}
         \caption{\label{pic:pdf_cpdBeforeCorrection_test} Geant4
           electronic recoil energy depositions in LAr for simulated
           background components scaled to counts per chain (multiple isotopes in the chain enter a component) or to
           counts per emitted radiogenic neutron in case of the \gray\ component
           ``neutron PMT glass''.}
\end{figure*}

The fit uses the Bayesian Analysis Toolkit
(BAT)~\cite{Caldwell:2009kh}, which uses a Markov Chain Monte Carlo to extract
posterior probability distributions based on prior probabilities and a user-defined likelihood function.

The complete posterior probability $p(\boldsymbol{\lambda} |
\mathbf{n})$ for the model with the set of parameters
$\boldsymbol{\lambda}$ given the data $\mathbf{n}$ is obtained with
Bayes' Theorem connecting the likelihood $p(\mathbf{n} |
\boldsymbol{\lambda})$ with the prior information for each parameter
$p_0(\boldsymbol{\lambda})$: 

\begin{equation}
 p(\boldsymbol{\lambda} | \mathbf{n}) = \frac{p(\mathbf{n} | \boldsymbol{\lambda}) \cdot p_0(\boldsymbol{\lambda})}{\int p(\mathbf{n} | \boldsymbol{\lambda}) \cdot p_0(\boldsymbol{\lambda}) \cdot d\boldsymbol{\lambda}}.
 \end{equation}

The denominator normalizes the probability.
The likelihood is expressed as the product of Gaussian
probabilities in each 200~PE$_{\rm corr}$ bin $i$ to observe $n_i$ events, given the
expectation $\mu_i$, based on the model parameters
$\boldsymbol{\lambda}$, and its uncertainty $\sigma_i$:

\begin{equation}
p(\mathbf{n} | \boldsymbol{\lambda}) = \prod_{{\rm bin}\, i} p(n_i | \boldsymbol{\lambda}) =  \prod_{{\rm bin}\, i}  \frac{1}{2\pi\sigma_i^2} \cdot \exp{\frac{(n_i-\mu_i)^2}{2\sigma_i^2}}. 
\end{equation}

$\mu_i$ is expressed as the sum of Monte Carlo expectations from each background component $k$ in bin $i$, $P_k({\rm PE}_i)$, scaled with the activity of the component $A_k$ and multiplied with a global scale parameter $S$ and its variation $\epsilon$: 

\begin{equation}
\mu_i = S \cdot \epsilon \cdot \sum_k A_k\cdot P_k({\rm PE}_i).
\end{equation}

$\epsilon$ reflects the uncertainties in cut acceptance.
$\sigma_i$ is the uncertainty in bin $i$ given by three components:

\begin{equation}
\sigma_i^2 = n_i +  \left( \sum_{k} w_{i,k} \sum_{l_k} w_{l_k} \cdot N_{i,l_k}^{\rm MC}\right)^2  + \left( n_i \cdot \sigma_{\rm bin} \right)^2.
\label{eq:likelihood_uncertainty}
\end{equation}

\noindent 
The first is the statistical uncertainty of the expected number of events
$n_i$.
The second is the sum over all statistical uncertainties in the simulated
distributions $N_{i,l_k}^{\rm MC}$ weighted by their component contribution
in the $i$-th bin $w_{i,k}$ and their nuclide contribution in the
background component $w_{l_k}$. 
For example, the \nuc{Tl}{208} Monte Carlo spectrum in the
\nuc{Th}{232} PMT component has a fixed $w_{l_k}=0.3594$ given by its
branching ratio in the decay chain and a $w_{i,k}$ which depends on
the \nuc{Th}{232} PMT contribution in bin $i$. 
The third component to $\sigma_i$ is an ad-hoc
uncertainty of $\sigma_{\rm bin}=0.03$ in each bin 
in order to stabilize the fit against small systematic effects in the
energy calibration. The large number of events in the data and Monte Carlo
simulations would require a highly accurate model (e.g.\ with sub
percent precision in each bin at the 2614.5~keV peak). The allowed bin-by-bin 
variation of 3\% was found to mitigate this problem while not
artificially inflating the posterior distributions.

The prior information for the activity parameters is based on
screening results or literature values and is listed further below
together with the fit results in \tab \ref{tab:PriorsAndPosteriors}. 
Three types of priors are distinguished
(1) a Gaussian prior for measurements in which the mean of the Gaussian
is the measured central activity value and the width is the 1$\sigma$
activity uncertainty, (2) the upper half of a Gaussian in cases where
only activity limits are known and (3) a flat prior for unbiased
measurements or where no prior knowledge is available.

The $k$ (free) activity parameters and the uncertainty on the global scale parameter $\epsilon$ span a
$k+1$-dimensional parameter space. 
A prior probability of 0 is assigned to all negative activity values.

The fit is performed in multiple stages to facilitate convergence and
reduce the required computational time. The Markov Chain Monte Carlo adopted in
BAT requires many sampling points in order to adequately
map the posterior probability space. The dimensional complexity of the
posterior increases with additional free parameters, so that a combined
fit of all activity components and energy response parameters was not
feasible. Furthermore, when energy response parameters are included as
free parameters, each simulated distribution has to be rebinned, rescaled, and
again smeared with a resolution response at each sampling point. This process tremendously increases the required CPU time.

The general strategy for fitting is to treat the energy response as
effective parameters and focus on extracting posterior activities
based on their prior knowledge. Thus, the sequence is as follows: (1)
Initially match the energy response parameters manually to data; (2)
Perform a preliminary fit with fixed energy response parameters to
obtain a good match of Monte Carlo simulations to data; (3) Fix the activity components from
(2) and perform a full fit of energy response parameters; (4) Fix the
energy response parameters from (3) and perform a component activity
fit w.r.t.\ the original prior knowledge. This approach does not allow
the calculation of parameter correlations between energy response and component
activities. However, the energy scaling to PE$_{\rm corr}$ as
described above does not allow a meaningful interpretation of the
energy response parameters to begin with. 
For the energy calibration in step (3), an additional free linear scaling parameter is allowed for the \nuc{Ar}{39} component which removes tension in the fit due to its dominant number of counts and its large Q-value uncertainty of 5~keV (0.9\%). The fit finds a 0.4\% shift in energy for \nuc{Ar}{39} w.r.t.\ the other components.
In the following, only step (4)
is discussed and later different variations of these fits are
performed to investigate systematic effects and different model
assumptions. The fit is performed from 2000 to 35000 in the linearized \PEC\ unit.

The global maximum of the extracted posterior distribution is the best
fit value for each parameter. Furthermore, the posterior distributions
are marginalized w.r.t.\ each parameter in order to obtain the smallest
connected 68\% credibility intervals (CI), which can be interpreted as
individual 1$\sigma$ parameter uncertainties. Finally,
two-parameter-pair distributions are marginalized to understand
parameter correlations.

%% file: results.tex
\begin{figure*}[h]
        \centering
                \includegraphics[width=0.95\textwidth]{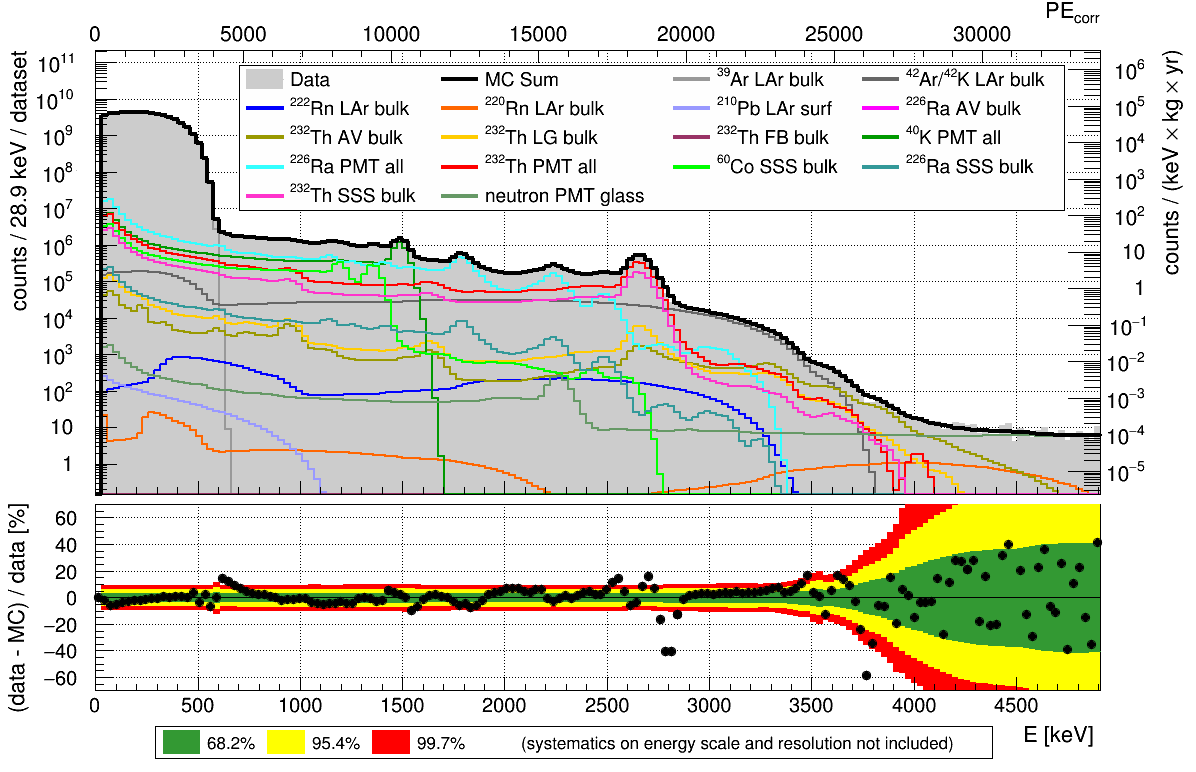}
         \caption{\label{pic:pdf_Spectrum_fullCompFit} Top panel: The
           energy spectrum of the ER data (shaded gray) is plotted
           with the fit result with the individual model components
           shown in thin colored lines and their sum in bold
           black. The linearized corrected PE variable is shown on the
           top axis. Bottom panel: Residuals of data and MC model in
           percent. Green, yellow, and red belts denote 1, 2, and 3
           $\sigma$, confidence, respectively, according to the
           uncertainty used in the likelihood definition. Systematics effects on the energy scale and resolution are not included as explained in the text.}
\end{figure*}

The energy spectrum of the dataset and the fit results are shown in
\fig \ref{pic:pdf_Spectrum_fullCompFit}, along with the fit residuals.
The ER background model describes the data well over 9 orders of
magnitude and a wide energy range. 
The residual shows a few outlying bins. 
The confidence belts in the residual plot are based on the uncertainties used in the likelihood (\eq \ref{eq:likelihood_uncertainty}) which do not include systematic effects from the energy scale and resolution. 
Discrepancies seen around steeply falling distributions at the \nuc{Ar}{39} endpoint (565 keV), the  \nuc{Tl}{208} peak (2615 keV), and the summation peaks around 3500 and 3750 keV are created by
inaccuracies in the energy response model. These do not significantly influence the derived activity of the components. 

\tab
\ref{tab:PriorsAndPosteriors} gives the prior information and the fit
results for each background component. The third column gives the
global maximum i.e.\ the best fit value and the last column gives the central 68\%
credibility interval of the marginalized posterior distributions.

\begin{table*}
\begin{center}
\caption{Bayesian fit results for activities are shown with the input
  priors. 
  Columns denote from left to right: 
  nuclides, decay chain or
  subchain included in the model; 
  prior information used in the fit;
  best fit results from the global maximum; 
  central connected 68\%
  credibility interval of the marginalized parameter space or the 90\%
  quantile in case of a limit. 
  The input can be Gaussian prior distributions (denoted by mean $\pm$ standard deviation), 
  a flat distribution (denoted with $[\ ]$ brackets),
  or a fixed value (denoted with $=$ ). 
  All priors are defined to be non-negative (i.e.\ Gaussian priors centered around
  zero denote upper activity limits). 
  The entry ``neutron PMT glass'' refers to
  the neutron production rate (in Hz) in a model where all neutrons come from the PMT glass. See text for a discussion of
  systematic uncertainties in the PMT glass.}
\label{tab:PriorsAndPosteriors}
\begin{ruledtabular}
\begin{tabular}{llllll}
Component   &   Input prior [Bq]           &   Best fit [Bq]        & Central 68\% interval [Bq]          \\
 \colrule
\nuc{Ar}{39} LAr bulk & $3282 \pm 340$ & $3009$ & $[2977 - 3042]$  \\ 
\nuc{Ar}{42}/\nuc{K}{42} LAr bulk & $[0-0.3]$ & $0.129$ & $[0.126 - 0.131]$  \\ 
\nuc{Rn}{222} LAr bulk & $=5.9\times 10^{-4}$ & - & -  \\ 
\nuc{Rn}{220} LAr bulk & $(8.5\pm 4.9)\times 10^{-6}$ & $7.4\times 10^{-6}$ & $<13.7\times 10^{-6}$  \\
\colrule 
\nuc{Pb}{210} LAr surf & $=2.0\times 10^{-4}$ & - & -  \\ 
\colrule
\nuc{Ra}{226} AV bulk & $(0 \pm 8)\times 10^{-2}$ & $0$ & $<3.9\times 10^{-2}$ (90\% CI)  \\ 
\nuc{Th}{232} AV bulk & $(0 \pm 22)\times 10^{-2}$ & $1.5\times 10^{-2}$ & $[1.1-1.6]\times 10^{-2}$  \\ 
\colrule
\nuc{Th}{232} LG bulk & $0 \pm 1.3$ & $0.13$ & $<0.2$ (90\% CI)  \\ 
\nuc{Th}{232} FB bulk & $0 \pm 0.9$ & $0$ & $<0.27$ (90\% CI)  \\ 
\colrule
\nuc{K}{40} PMT all & $[500-1500]$ & $776$ & $[757 - 795]$  \\ 
\nuc{Ra}{226} PMT all & $216 \pm 24$ & $136$ & $[131 - 137]$  \\ 
\nuc{Th}{232} PMT all & $39 \pm 4$ & $41.5$ & $[38.1 - 44.4]$  \\ 
neutron PMT glass & $[0-5]\times 10^{-2}$ & $1.47\times 10^{-2}$ & $[1.33 - 1.62]\times 10^{-2}$  \\ 
\colrule
\nuc{Co}{60} SSS bulk & $78 \pm 11$ & $45.0$ & $[42.5 - 47.5]$  \\ 
\nuc{Ra}{226} SSS bulk & $10.6 \pm 5.8$ & $4.9$ & $<12.9$ (90\% CI)  \\ 
\nuc{Th}{232} SSS bulk & $9.7 \pm 5.6$ & $43.0$ & $[36.9 - 49.0]$  \\ 
\end{tabular} 
\end{ruledtabular}
\end{center}
\end{table*}

\begin{figure*}[h]
        \centering
                \includegraphics[height=0.8\textheight]{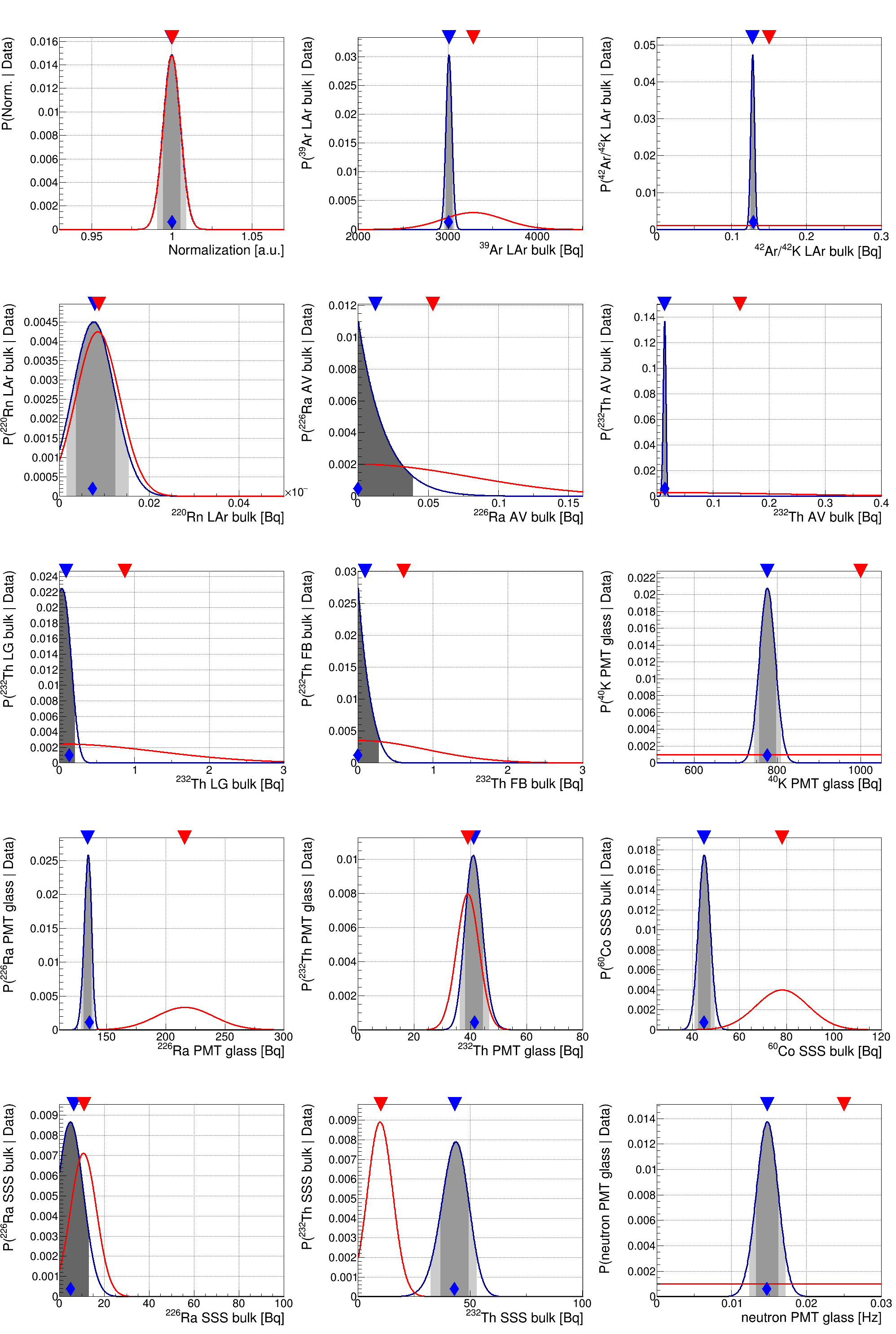}
         \caption{\label{pic:pdf_compArray_fullFit} Priors (red) and
           marginalized posterior distributions (blue) of free fit
           parameters. The blue diamond denotes the global best fit
           value for this parameter. The red and blue triangles denote
           the median of the prior and posterior distribution,
           respectively. In case of a posterior probability distinct
           from zero by 2$\sigma$, the gray shaded regions show the 1
           and 2$\sigma$ credibility intervals, respectively and
           otherwise the 90\% quantile is drawn in dark gray.  Note
           that the whole parameter range is not plotted; for flat
           priors the median indicates the center of that range.}
\end{figure*}

\begin{figure*}[h]
        \centering
                \includegraphics[width=0.99\textwidth]{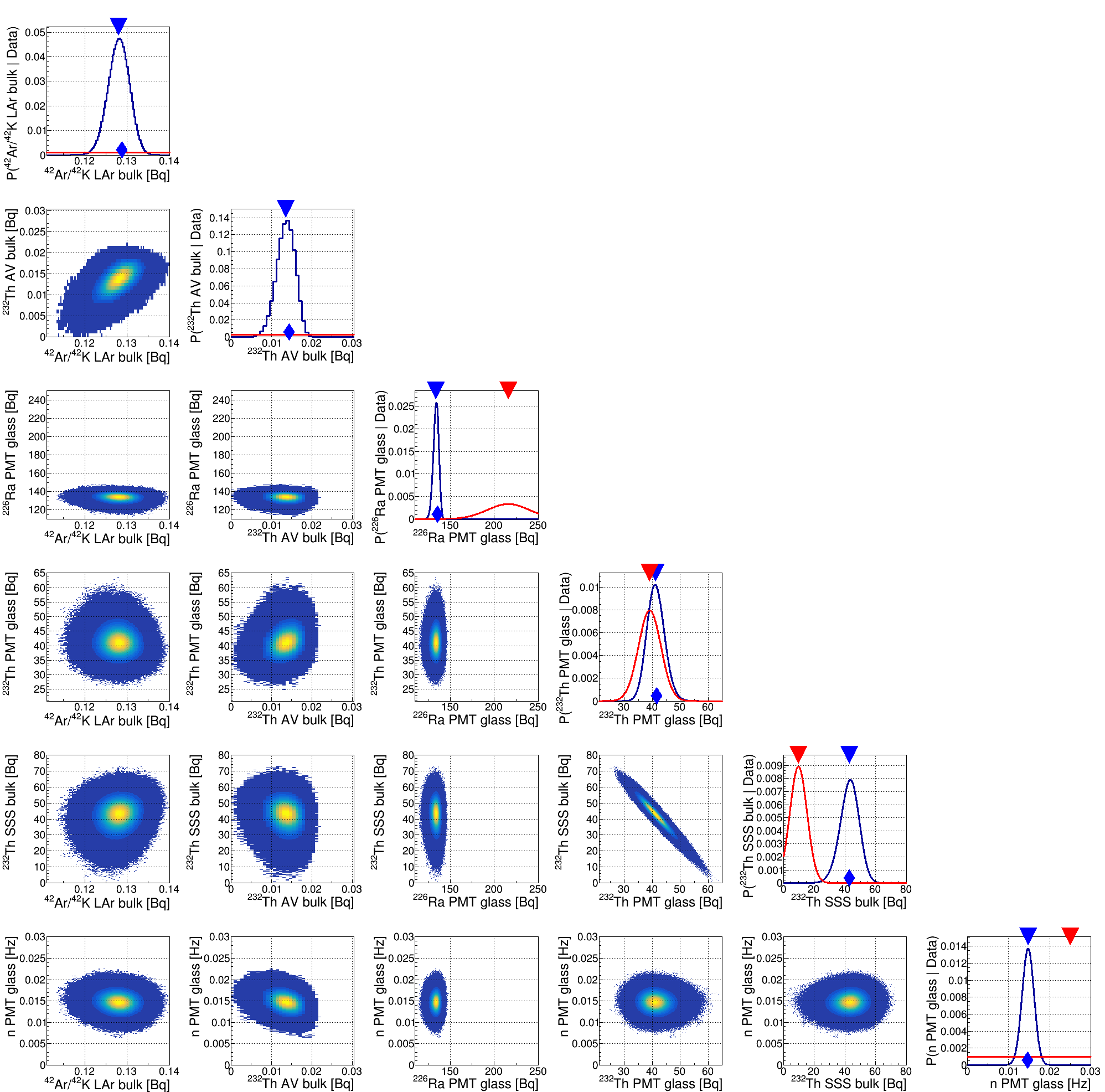}
         \caption{\label{pic:pdf_corrArray_fullFit} Correlations
           between selected components in global fit. The scatter
           plots show the marginalized two-parameter posterior
           activity distribution in Bq in which the color scale denotes the
           probability. The one
           dimensional plots show the prior distribution in red and
           the marginalized single-parameter posterior distribution in
           blue. The colored triangles show the median of these
           distributions, respectively and the diamond shows the
           global maxium. }
\end{figure*}

The prior distributions entering the fit and the marginalized
posterior distributions are shown in \fig \ref{pic:pdf_compArray_fullFit}.  The two-parameter pair posterior
distributions are shown in \fig \ref{pic:pdf_corrArray_fullFit} for
selected components and are discussed in detail below.
 
The first parameter in \fig \ref{pic:pdf_compArray_fullFit} is
the global scale parameter uncertainty $\epsilon$ with a Gaussian prior centered around 1 with the width 
as the uncertainty of cut efficiencies. The posterior
distribution follows the prior distribution and thus no knowledge
update is obtained from the fit, but uncertainty correlations are naturally included in the activity results.

The \nuc{Ar}{39} activity has a prior probability distribution peaked at 3282~Bq with 10\% uncertainty, 
from \cite{Ar39_warp}. After the fit, the posterior probability distribution is found to have a peak at 3009~Bq with a 1\% uncertainty
 or in a 68\% credibility interval of
     [$2977-3042$]~Bq which is about 10 times more precise than
     in \cite{Ar39_warp}. However, since the fit is focused on the
     higher energy parts of the ER band starting at 290~keV and the
     residual plot shows unaccounted energy shape systematics in
     the \nuc{Ar}{39} region, at this time we refrain from quoting this number as a
     new specific activity measurement in atmospheric LAr and defer to
     a dedicated future analysis.

A flat prior is used for the \nuc{Ar}{42}/\nuc{K}{42} activity in
order to not rely on debated values in the literature. The posterior
distribution has a width of about 2\%, but excludes important
systematics from the energy scale and the $\beta$-decay spectral
shape. A dedicated analysis is presented in \sec
\ref{sec:Ar42K42}.

\nuc{Rn}{222} and its short-lived daughters in the LAr bulk, as
well as the \nuc{Pb}{210} contribution on the LAr-TPB interface, are
included in the model but fixed in the fit. Their activities were
determined from dedicated $\alpha$-decay analyses~\cite{PRL} and are
negligible in the ER band. These components serve mainly an
illustrational purpose.  Internal \nuc{Rn}{220} and its daughter
\nuc{Tl}{208} can create the highest energy radioactive decay signals  in DEAP-3600 with energies up to its Q-value of 4999.0~keV. 
The spectrum is dominated by two $\beta$ shape contributions, as shown in \fig \ref{pic:pdf_Spectrum_fullCompFit}: one where the 2614.5~keV
\gray\ escapes the LAr and one where it is contained. However, at its
highest significance at around 4.3~MeV, 
this component is more than an order of magnitude lower than the signal from \grays\ produced by neutron capture reactions.
Thus, the sensitivity to \nuc{Rn}{220} in the LAr bulk
is small, and the posterior follows the prior information from the
$\alpha$-decay analysis as shown in \fig \ref{pic:pdf_compArray_fullFit}.

The external \nuc{Th}{232} chain contribution with the 2614.5~keV
\gline\ as its most dominant feature is shared between the close
sources in the AV, LG and FB bulks as well as between far sources in
the PMTs and the SSS bulk. All of these components are left floating in
the fit, with prior constraints.  The main discrimination power between
sources close to and far from the LAr 
comes from the summation peaks.  For the sources close to the LAr, 
only upper limits are available as input prior. The fit finds a
significant source in the AV bulk at $15$~mBq, compared to the
previously known limit of $<220$~mBq. The LG contribution is
consistent with 0, but the best fit value is found at 134~mBq (see \fig \ref{pic:pdf_Spectrum_fullCompFit} and
\ref{pic:pdf_compArray_fullFit}).  No significant correlation of close
sources (represented by the AV bulk concentration in \fig \ref{pic:pdf_corrArray_fullFit}) with far sources (PMT, SSS) is observed. 
The activity of \nuc{Th}{232} in the sources far from the LAr is found to be significantly higher than expected, with strong correlations between the best fit activities in these sources.
This additional \nuc{Th}{232} activity is mainly put into the SSS bulk due its less
precise screening results and therefore wider prior
distribution. 

For dark matter experiments such as DEAP-3600, it is important to
understand uncertainties on the activities in PMTs, as ($\alpha$,n)
reactions in borosilicate glass are a source of fast neutrons. Furthermore, the prior estimates for both the PMT
components and the stainless steel in the fit were based on sample
measurements which may not be completely representative of the
whole. To address these concerns the fit was repeated twice: first,
the contamination from \nuc{Th}{232} in the stainless steel was fixed
at its prior of 9.7 Bq and the contamination from PMT glass was
allowed to float, resulting in an activity of 58.0 Bq. Second, the
activity of \nuc{Th}{232} in the PMTs was fixed at the prior value of
39.0 Bq and the contamination in the SSS was allowed to float,
resulting in a value of 70.6 Bq. Including this systematic effect, the
activity of the \nuc{Th}{232} in the PMT glass is within the range of 38 to 58~Bq.

The dominant presence of \nuc{Ra}{226} (subchain in the \nuc{U}{238} decay chain) is expected
to be in the PMTs and the SSS bulk. The most
prominent features in the ER spectrum are the peaks from the
1764.5~keV and 2204.1~keV \grays\ from \nuc{Bi}{214}. These features
are dominated by the PMT's contribution by about 2 orders of
magnitude. For the AV bulk contribution only prior limits are known. 
The fit finds 0 as best fit value while being able to constrain the
limit further. The PMT's contribution is found to be a factor of 1.6 lower
than the prior expectation which is about 3.3~$\sigma$ lower than {\it ex-situ} assay values. 
This indicates a non-representative
screening sample or a misleading correlation with the less well-known
SSS bulk activity. No strong correlation is observed with other background components, indicating that the sensitivity to
this component is indeed coming from the \glines\ and not from the
continuum in the spectrum.

The single \gray\ from \nuc{K}{40} does not allow disentangling activities of
components at different distances, i.e.\ from the AV, LGs, FBs, and the
PMTs. Without an assumption of the source positions, the screening
knowledge cannot be converted into a prior activity and thus a flat prior was chosen. 
All \nuc{K}{40} events in the spectrum are interpreted as coming from the PMTs, which are expected to have the highest 
activity. An effective activity of 776~Bq is determined, compared to
an expected activity of 454~Bq. The discrepancy could be explained by
lower activity sources closer to the target, i.e.\ the acrylic and PE
material in the AV, LG and FB.

The \nuc{Co}{60} prior activity from screening the SSS is known with
15\% precision. The fit finds a factor of 1.7 lower activity than
expected. This could be due to a non-representative screening
sample.

The component of \grays\ from neutron capture reactions is included into the fit with a flat
prior. Their contribution is dominant above about 4~MeV,
which allows a precise measurement without strong correlations with
other sources in the fit. The \gline\ from neutron capture
on \nuc{H}{1} at 2224.6~keV is clearly visible in the simulated spectrum, but
subdominant compared to other signals in this region. \\

Based on the full model, two dedicated fits were performed to investigate the hypothetical contributions of emanated \nuc{Rn}{220} and \nuc{Rn}{222} within the steel shell freezing out on the outer AV surface (see \nuc{Rn}{220} RnEm and \nuc{Rn}{222} RnEm in \fig \ref{pic:BackgroundComponents}). In a first fit, these component were added to the full model. The experimental signature is dominated by \nuc{Tl}{208} and \nuc{Bi}{214}, respectively, which differ only slightly in their summation peak ratios compared to other close components. The fit finds zero activity as best fit values, however, strong correlations with other close sources are observed. As a worst case scenario, a second fit without AV, LG, FB components was performed so that the \nuc{Rn}{220} eman and \nuc{Rn}{222} eman components account for all summation peak features. This fit finds $0.019\pm0.001$~Bq and $<0.06$~Bq for \nuc{Rn}{220} and \nuc{Rn}{222}, respectively, which constrains the activity of radon daughters for ($\alpha$,n) production close to the LAr target to $<0.02$~Bq and $<0.06$~Bq, respectively.

%% file: externalgammaverification.tex
The analysis chain and Monte Carlo simulations were tested with a \nuc{U}{232} calibration source in a
source deployment tube just outside the steel shell. 
\nuc{U}{232} (\THL=$68.9\pm0.4$~yr) decays into \nuc{Th}{228}, emulating the lower part of a
distant \nuc{Th}{232} decay chain as might be seen from contributions in the SSS or the PMTs. 
The total trigger rate is only increased by 1.4\% and does not significantly change the detector system during these calibration campaigns.

Only \nuc{Tl}{208} decays are simulated in the source since
\grays\ from other \nuc{U}{232} decay chain isotopes are strongly
attenuated. The calibrated \nuc{U}{232} source activity was $15.3\pm1.5$~kBq leading to a \nuc{Tl}{208} activity of $5.51\pm0.55$~kBq.  
A 22.16~h source run was taken and analyzed using the same procedure as used for the standard dataset. 
The full ER background model was used, with all components, except for \nuc{Ar}{39}, fixed to their best fit values.
The activity of the \nuc{U}{232} source was given a flat prior and floated in the fit.


The best fit value for the \nuc{U}{232} source activity is 19\% lower than expected from the datasheet of the source, though we point out that the activity stated on the datasheet has an uncertainty of 10\%. The 68\% confidence interval from the fit corresponds to 0.7\% of the fit value, which does not include systematic uncertainties from positioning and missing details in the Geant4 model of the steel shell. Small variations of the amount of water as well as screws and bolts can change the attenuation of \grays\ in-between the source and the LAr.
Such uncertainties are expected to be much smaller for background components within the SSS whose positions are well known and have been carefully verified. We thus consider this a satisfactory validation of the model.

%% file: Ar42K42.tex

\nuc{Ar}{42} is present as a cosmogenic trace isotope in atmospheric
argon. The production mechanisms are (1) cosmic $\alpha$-particles
inducing ($\alpha$,2p) reactions on \nuc{Ar}{40} in the outer
atmosphere and (2) double neutron capture
\nuc{Ar}{40}(n,$\gamma$)\nuc{Ar}{41}(n,$\gamma$)\nuc{Ar}{42}.  The
production by process (1) is estimated to be dominant and constant over
time with a \nuc{Ar}{42}/\nuc{Ar}{40} ratio of $\approx 10^{-20}$~g/g
\cite{Ar42Prod1}, whereas the production via process (2) requires the
high neutron flux from atmospheric nuclear weapon testing, which
stopped in the 1980's and can thus only account for a \nuc{Ar}{42}/\nuc{Ar}{40} ratio of
$\approx 10^{-22} - 10^{-21}$~g/g \cite{Ar42Prod2}.

Due to its long half-life of \THL=$32.9\pm1.1$~yr, \nuc{Ar}{42} is
well-mixed in the atmosphere, and its specific activity today is
constant in time. The short half-life of its daughter \nuc{K}{42}
(\THL=12.4~h) means that the isotopes are in secular equilibrium. 
In DEAP-3600, \nuc{K}{42} $\beta$-decays with
$3525.2\pm0.2$~keV endpoint are easier to measure than \nuc{Ar}{42}
$\beta$-decays with $599\pm6$~keV endpoint.

\begin{table*}
\begin{center}
\caption{\label{tab:K42Literature} Previous results on specific activities of \nuc{Ar}{42} in atmospheric LAr. The various detection techniques are indicated with a simplified description.}
\begin{ruledtabular}
\begin{tabular}{llll}
Reference  &  Year  &  Technique  & \multicolumn{1}{c}{Specific activity [$\mu$Bq/kg]}\\
\hline
Ashitkov et al. \cite{Ashitkov98}      &  1998     & LAr ion.\ det.                              &          $<61.4$ (90\% CL) \\
Ashitkov et al. \cite{Ashitkov03}      &  2003     & LAr ion.\ det.                               &         $<44.0$ (90\% CL)  \\
GERDA collaboration \cite{Gerda16}      &  2016     & HPGe $\gamma$-spec.             &          $=91^{+8}_{-20}$ - $168^{+22}_{-18}$   \\
Barabash et al. \cite{Barabash16}      &  2016     & LAr ion.\ det.                               &          $=92^{+22}_{-46}$   \\
This work & 2019 & scintillation &  $= 40.4 \pm 5.9 $\\


\end{tabular}
\end{ruledtabular}
\end{center}
\end{table*} 


The concentration of \nuc{Ar}{42} in atmospheric argon was measured in the past.
Ashitkov et al.\ studied \nuc{K}{42} decays with a
LAr ionization detector \cite{Ashitkov98,Ashitkov03}. These data were
re-analyzed by Barabash et al.\ \cite{Barabash16}. The GERDA experiment
\cite{Gerda16} used HPGe detectors immersed in a large LAr tank to
measure the specific activity. 
The different measurements of the \nuc{K}{42} activity, collated in \tab
\ref{tab:K42Literature}, are in tension with each other.
For the LAr ionization detector \cite{Barabash16}, the energy region
between 3.1 and 3.6~MeV is used to count events coming from $\beta$'s in
the tail end of the \nuc{K}{42} spectrum, while subtracting events
expected from backgrounds. An activity of $92^{+22}_{-46}$~$\mu$Bq/kg
is quoted where the background subtraction and energy calibration of
this region are the largest sources of systematic uncertainties.

In the GERDA experiment, the measured activity is extracted from a full
background model fit, mainly sensitive to the 1524.6~keV peak counts
from the \nuc{K}{42} excited state transition which has an 18\% branching ratio. 
Different model assumptions with different complexity yield
results varying between $91^{+8}_{-20}$ and $168^{+22}_{-18}$
$\mu$Bq/kg \cite{Gerda16}.  The \nuc{Ar}{42} concentration in LAr can
be assumed homogeneous, but GERDA observed that a fraction of the
\nuc{K}{42} daughters is charged after the initial \nuc{Ar}{42} decay,
and its concentration can be distorted by electric fields. Especially
in the case of GERDA with high voltage biased germanium detectors, this leads
to the attraction of \nuc{K}{42} ions towards the n+ surface of the
detectors at positive 2 to 4~keV \cite{Gerda16}. The issue of
\nuc{K}{42} ion drift was mitigated with a copper shroud around the
detector strings, shielding the electric field, and the above activities are obtained assuming a homogenous \nuc{K}{42} distribution. 

In DEAP-3600 no electric field is present 
and the \nuc{Ar}{42} and \nuc{K}{42} concentrations can be assumed
homogeneous. The activity is extracted from the full ER background model
fit, which is most sensitive in the region above 2.8 MeV. Thus
DEAP-3600 provides a third independent measurement with
different systematic uncertainties which are explored below.

The best fit of the total \nuc{Ar}{42}/\nuc{K}{42} activity in the DEAP-3600 LAr is
$0.129$~Bq, which translates into a specific activity of
39.6~$\mu$Bq/kg. The marginalized posterior distribution yields a
width of about 2\% uncertainty. However, systematic uncertainties dominate as follows.

The decay of \nuc{K}{42} results in a single $\beta$ with an endpoint of
3.525~MeV, with 81.9\% branching ratio. Thus it has a different topology, and
different saturation effects, than the multiple \gray\ interactions of
\nuc{Tl}{208} sources close to the LAr, which is the dominant background for this measurement
above 2.8~MeV. To estimate the uncertainty from the energy scale, the
analysis was repeated without the topology corrections, resulting in a
fit activity of 0.113~Bq, or a 13\% decrease. This is taken as the conservative
systematic uncertainty from energy scale uncertainties.

The \nuc{K}{42} decay is expected to be isotropic in the detector and
will thus have different saturation effects than external single
\grays\ which interact close to the edge of the detector. In order to
estimate this effect, the default energy scale is used, except for the
\nuc{K}{42} component, which is assigned a free linear energy scaling
parameter. The energy scale correction parameter for \nuc{K}{42} is
fitted between 20000 and 34000~\PEC\ simultaneously with the
\nuc{K}{42} activity and the activity of the relevant contributions in
this range. In this case, the energy scale of \nuc{K}{42} is increased by 0.32\% and the change in derived activity
of \nuc{K}{42} is less than 0.8\%.

The \nuc{K}{42} $\beta$-decay into the ground state of \nuc{Ca}{42} is
a $2^-$ to $0^+$ 1st-forbidden unique transition with theoretical
uncertainties on the spectral shape. The analysis was repeated with an
allowed $\beta$ spectrum used instead of the 1st-forbidden spectrum. The
best fit value changes to 0.123~Bq ([$0.120 - 0.125$]~Bq in a 68\% CI). We
conservatively take this 4.7\% shift as the systematic uncertainty
from uncertainties in the spectral shape.

Three additional uncertainties are considered which directly affect the activity estimate: 
(1) uncertainty on the LAr mass in the detector, 
(2) uncertainty on the MC simulation, and 
(3) uncertainty on the age of the argon, measured from the time the argon was extracted from the atmosphere.
The time from atmospheric extraction
by the vendor to the start of data taking is estimated as 0.5~yr but has a large uncertainty. Thus,
the average age during the 1~yr dataset is about $1.0\pm0.5$~yr which
reduces the activity by $2.1\pm1.0$\% with respect to atmospheric argon.  All systematic uncertainties
are summarized in \tab \ref{tab:SystematicsK42}.


\begin{table}[ht]
\begin{center}
\caption{Systematic uncertainties for \nuc{Ar}{42}/\nuc{K}{42} activity measurement.}
\label{tab:SystematicsK42}
\begin{ruledtabular}
\begin{tabular}{lc}
\toprule
Systematics   &    Fraction of activity    \\
 \hline
Fit uncertainty          & 2\%  \\ 
MC simulation          & 3\%  \\ 
LAr mass                 & 3.4\%  \\ 
Nuclear physics       & 4.7\%  \\ 
Energy scale           &  $<0.8$\%  \\ 
Topology correction & 13\%  \\ 
\hline
Subtotal                   & 14.7\%\\
\hline
Age of LAr &  1\%  \\ 
\bottomrule
\end{tabular} 
\end{ruledtabular}
\end{center}
\end{table} 

The measured specific activity of \nuc{Ar}{42}/\nuc{K}{42} for LAr in
DEAP-3600 is $39.6 \pm 5.8 $~$\mu$Bq/kg. 
Extrapolating back to the time of extraction from the atmosphere, the specific activity is estimated to be
\begin{equation}
\rm A = 40.4 \pm 5.9~\mu Bq/kg.
\end{equation}


This value is significantly lower than found in previous experiments
\cite{Gerda16,Barabash16}. 
 

%% file: conclusion.tex
Electromagnetic backgrounds are described for the DEAP-3600
detector. They are modeled over an energy range from 290~keV to 5~MeV
and more than 9 orders of magnitude in vertical scaling. The activity
of \nuc{Ar}{42}\ in atmospheric argon has been found to be $40.4 \pm
5.9 \mu\mbox{Bq/kg}$ which is 
lower than measured in previous work
and represents a significant reduction in uncertainty.

Of particular importance to the WIMP search, the activity in the PMT
glass has been measured. In the Bayesian model, the best fit
for \nuc{Th}{232} activity was 41.5~Bq (68\% statistical credibility
interval between 38.1 and 44.4~Bq) and between 38 and 58~Bq after including
systematic effects from correlations with the activity in the
stainless steel shell. The activity of the \nuc{Ra}{226} subchain in the \nuc{U}{238} decay chain was found to be
136~Bq (68\% statistical credibility interval between 131 and 137~Bq). 

In addition, the hypothetical contribution of emanated \nuc{Rn}{220} and \nuc{Rn}{222} plated out on the outer AV surface could be constrained to $<0.02$~Bq and $<0.06$~Bq, respectively, for a worst-case scenario.

%% file: appendix.tex
\subsection{Material Assays}
\label{sec:MaterialAssays}

The specific radioactivities of the main materials are shown in \tab \ref{tab:ScreeningResults1} along with the number of components in DEAP-3600 and their total mass. 
The specific activities are quoted as 90\% upper limits for measurements below the assay sensitivity and with 1$\sigma$ uncertainties  otherwise. \\

\begin{table*}
\begin{center}
\caption{Screening results for dominant ER background components. Activities are reported with 1$\sigma$ uncertainties. A 90\% confidence limit is placed when the measurement is below the background sensitivity of the detector.}
\label{tab:ScreeningResults1}
\begin{ruledtabular}
\begin{tabular}{lllllll}

Component  & Volumes & Total mass & \nuc{Ra}{226}     & \nuc{Th}{232}       & \nuc{K}{40}          & \nuc{Co}{60} \\
	            &        &  [kg]   &  [mBq/kg]             & [mBq/kg]               & [mBq/kg]              & [mBq/kg] \\
 \hline
 acrylic vessel                                  & 1     &  $643\pm64$         & $<0.1$                  &  $<0.5$   &  $2.1 \pm 1.8$ & - \\
 light guide acrylic                            &  255 &  $3774\pm377$     & $<0.1$                &  $<0.3$                &  $<1.1$                & - \\
 filler block HDPE                             &  486  &  $2725\pm273$    & $0.4\pm0.3$     &  $<0.1$     &  $<5.4$   & - \\
304 stainless steel                           &  1     &  $5012\pm501$     & $2.1 \pm 1.1$   &   $1.9 \pm 1.1$  &  -  & $15.5 \pm 1.7$\\

\end{tabular}
\end{ruledtabular}
\end{center}
\end{table*}

Special care is taken for the PMTs which contain several materials with high specific activity. Individual PMT components were separated and measured individually  \cite{DEAPDetector,GammaSpectroscopyAssay}. The specific activities as well as total activity budget of the PMTs is shown in \tab \ref{tab:PMTScreening}.

\begin{table*}
\begin{center}
\caption{HPGe $\gamma$-spectroscopy screening results from screening and scaling of individual PMT components. The mass as well as the specific activity and component activity are shown for the three primordial isotopes. For the \nuc{U}{238}, the measured activity of the lower chain (below \nuc{Ra}{226}) is quoted. The uncertainty of the component activity and subtotals is the combination from mass and specific activity uncertainty.}
\label{tab:PMTScreening}
\begin{ruledtabular}
\begin{tabular}{lllll | lllll}

Component          & Mass  [g]   & \multicolumn{ 3}{c|}{Specific activity [mBq/kg]}    & \multicolumn{ 3}{c}{Component activity  [mBq]}     \\

		   &               & \nuc{Ra}{226} & \nuc{Th}{232} & \nuc{K}{40}    & \nuc{Ra}{226} & \nuc{Th}{232} & \nuc{K}{40}   \\
 \hline
	Glass &              $741\pm100$    &   $921 \pm 34$  &   $139 \pm 7$  &   $546 \pm 66$             &    $683 \pm 95$   & $103 \pm 15$   & $405 \pm 70$  \\
	Ceramic &         $25.3\pm2.5$    &   $979 \pm 56$  &   $245 \pm 28$  &   $13771 \pm 1300$    &    $24.8 \pm 2.9$   & $6.2 \pm 0.9$   & $348 \pm 48$  \\
	Feedthrough &  $56.1\pm5.6$    &   $1138 \pm 60$  &   $430 \pm 32$  &   $9497 \pm 931$       &    $63.8 \pm 7.2$   & $24.1 \pm 3.0$   & $533 \pm 75$  \\
	Metal &             $120\pm12$       &     $<5.0$  &   $<3.3$    &    $1148 \pm 152$                     &   $<0.6$  & $<0.4$   & $137 \pm 23$  \\
	PVC Mount &   $1080\pm108$    &   $72 \pm 5$  &   $18.6 \pm 2.5$  &   $329 \pm 48$             &    $77.5 \pm 9.3$   & $20.1 \pm 3.4$   & $356 \pm 63$  \\
\hline
	Total PMT & $2022\pm148$~g &   &  &    & $849 \pm 96$   & $153 \pm 15$  &  $1779 \pm 131$ \\
\hline
\hline
	Total DEAP & $516\pm38$~kg  &   &  &    & $216 \pm 24$~Bq   & $39 \pm 4$~Bq  &  $454 \pm 33$~Bq \\

\end{tabular}
\end{ruledtabular}
\end{center}
\end{table*}

Materials with smaller masses and smaller radioactivity were evaluated but not included into the model in order to reduce complexity. 
The \nuc{C}{14} content of the AV acrylic was assayed with Accelerator Mass Spectrometry at Ottawa University \cite{AMSOttawaU} to less than 0.022 Bq/kg C. This amounts to a negligible contribution of $<12$~Bq in the whole AV. 
Other stainless steel components used as connectors between LGs, FBs and PMTs account for about 250~kg of stainless steel. This amounts to a total of approximately 0.56~Bq of \nuc{Ra}{226}, 0.51~Bq of \nuc{Th}{232}, 1.3~Bq of \nuc{K}{40} and 4.1~Bq of \nuc{Co}{60}. These components are subdominant compared to the activity of the PMTs at about the same distance to the LAr and subdominant compared to the mass of the steel shell at a slightly larger distance.
 Other materials such as insulating polyurethane foam between PMTs and steel shell were assayed but found to be negligible with a total mass of about 80~kg.